\documentclass[aps,prc,twocolumn,groupedaddress]{revtex4-2}

\usepackage{graphicx}
\usepackage{slashed}
\usepackage{ulem}
\usepackage{color}
\newcommand{\nopieft}{\mbox{$\slashed{\pi}$EFT~}}

\begin{document}

\title{Consequences of increased hypertriton binding for $s$-shell $\Lambda$-hypernuclear systems}

\author{M. Sch\"{a}fer}
\email[]{schafer.martin@mail.huji.ac.il}
\affiliation{Racah Institute of Physics,The Hebrew University, Jerusalem 91904, Israel}
\affiliation{Nuclear Physics Institute of the Czech Academy of Sciences, 25068 \v{R}e\v{z}, Czech Republic}

\author{B. Bazak}
\email[]{betzalel.bazak@mail.huji.ac.il}
\affiliation{Racah Institute of Physics,The Hebrew University, Jerusalem 91904, Israel}

\author{N. Barnea}
\email[]{nir@phys.huji.ac.il}
\affiliation{Racah Institute of Physics,The Hebrew University, Jerusalem 91904, Israel}

\author{A. Gal}
\email[]{avragal@savion.huji.ac.il}
\affiliation{Racah Institute of Physics,The Hebrew University, Jerusalem 91904, Israel}

\author{J. Mare\v{s}}
\email[]{mares@ujf.cas.cz}
\affiliation{Nuclear Physics Institute of the Czech Academy of Sciences, 25068 \v{R}e\v{z}, Czech Republic}

\date{\today}

\begin{abstract}
Consequences of increasing the binding energy of the hypertriton ground 
state ${_{\Lambda}^3}{\rm H}(J^P={\frac{1}{2}}^+)$ from the emulsion value 
$B^{\rm EMUL}_{\Lambda}({_{\Lambda}^3}{\rm H}_{\rm g.s.})$=0.13$\pm$0.05 MeV to the STAR value $B^{\rm STAR}_{\Lambda}(^{3}_{\Lambda}{\rm H}) = (0.41\pm 0.12 \pm 0.11)$~MeV are studied 
for $s$-shell hypernuclei within a pionless EFT approach at leading order, 
constrained by the binding energies of the $0^+$ and $1^+$ ${_{\Lambda}^4}
{\rm H}$ states. The stochastic variational method is used in 
bound-state calculations, whereas the inverse analytic continuation in the 
coupling constant method is used to locate $S$-matrix poles of 
continuum states. It is found that the $\Lambda nn({\frac{1}{2}}^+)$ resonance 
becomes broader and less likely to be observed experimentally, whereas the 
${_{\Lambda}^3}{\rm H}({\frac{3}{2}}^+)$ spin-flip virtual state moves closer to the $\Lambda d$ threshold to become a shallow bound state for specific $\Lambda N$ interaction strengths.  The effect of  such a  near-threshold ${_{\Lambda}^3}{\rm H}({\frac{3}{2}}^+)$ 
state on femtoscopic studies of $\Lambda$-deuteron correlations, and its 
lifetime if bound, are discussed. Increasing $B_{\Lambda}({_{\Lambda}^3}
{\rm H}_{\rm g.s.})$ moderately, up to $\sim$0.5 MeV, hardly affects calculated values of
$B_{\Lambda}({_{\Lambda}^5}{\rm He})$.
\end{abstract}
\maketitle

\section{Introduction}
\begin{figure*}[t]
\includegraphics[width=\textwidth]{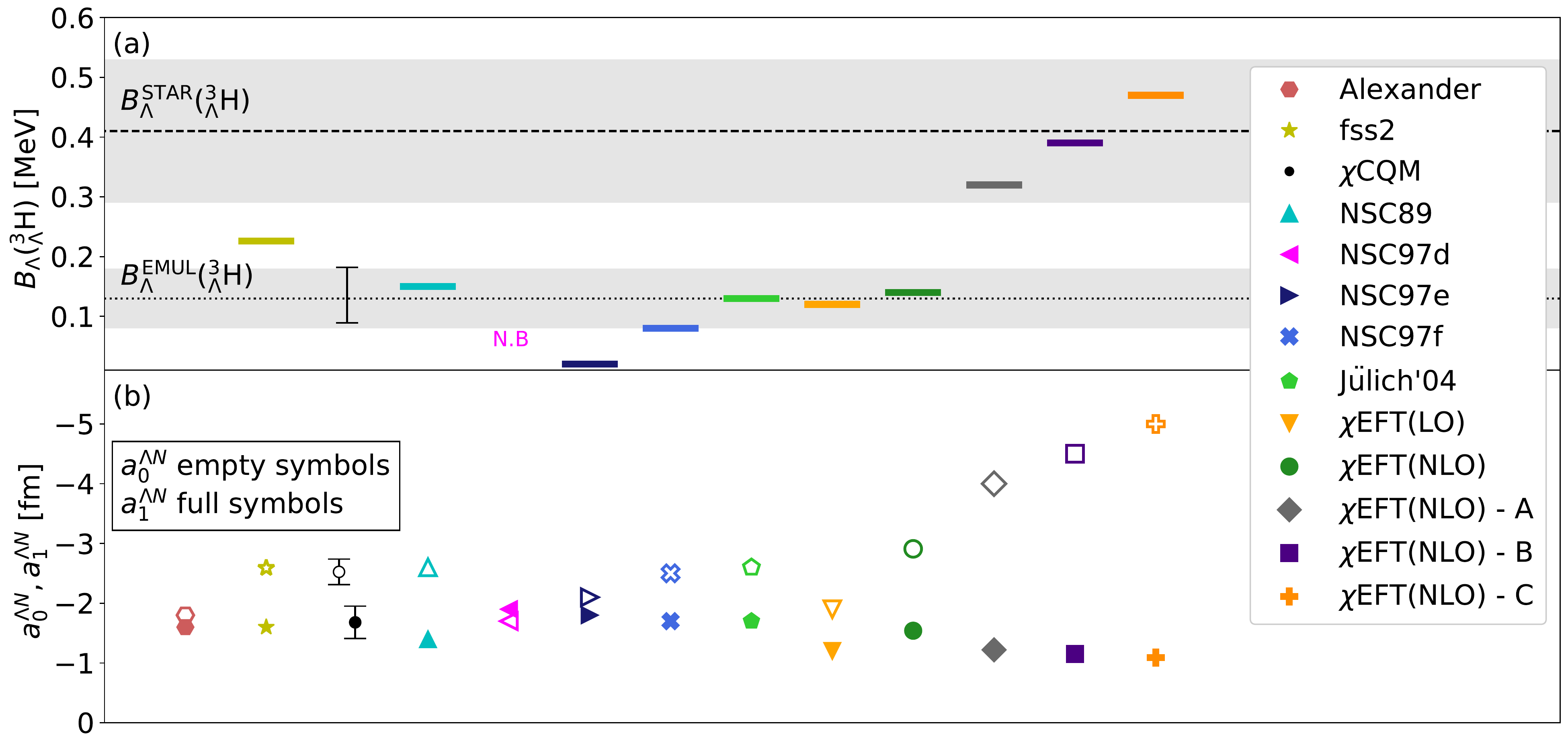}
\caption{The $\Lambda N$ spin-singlet ($a_0^{\Lambda N}$) and spin-triplet ($a_1^{\Lambda N}$) scattering lengths predicted by analysis of $\Lambda p$ scattering data (Alexander \cite{alexander68}) or by various $\Lambda N$ interaction models - fss2 \cite{FSKM08}, $\chi$CQM \cite{GFV07}, NSC89 \cite{NSC89}, NSC97(d,e,f) \cite{NSC97}, J\"{u}lich'04 \cite{julich04}, $\chi$EFT(LO) \cite{lo06}, $\chi$EFT(NLO) \cite{nlo13}, and $\chi$EFT(NLO) - A, B, C \cite{le20} (b). The upper panel (a) shows calculated $B_\Lambda({\rm ^3_\Lambda H})$ \cite{FSKM08,GFV07,nogga13,le20} for respective $\Lambda N$ interaction model compared to experimental values $B^{\rm EMUL}_\Lambda ({\rm ^3_\Lambda H})$ \cite{ju73} and $B^{\rm STAR}_\Lambda ({\rm ^3_\Lambda H})$ \cite{star20}. All $\chi$EFT results are for momentum cutoff $\lambda = 500~{\rm MeV}$.}
\label{interactions}
\end{figure*}

Interactions between hyperons and nucleons are not known in sufficient detail because the relevant scattering data are scarce, of limited accuracy and do not contain direct information on the spin dependence of the interactions. Moreover, scattering experiments do not fix directly the interactions at very low energies or even below threshold. 
Under such circumstances, one has to resort to bound few-body systems which thus serve as an important testing ground for the underlying baryon interactions. The hypertriton $^{3}_{\Lambda}$H ($J^{\pi} = 1/2^+$), being the lightest bound hypernucleus, holds a prominent position among these  systems (like the deutron in the case of the NN  interaction). The $\Lambda$ separation energy of $^{3}_{\Lambda}$H has been used as a constraint for various interaction models for decades. The widely accepted value $B^{\rm EMUL}_{\Lambda}(^{3}_{\Lambda}{\rm H}) = 0.13\pm 0.05$~MeV was extracted from four different sets of emulsion data~\cite{ju73}. The spin $S=1/2$ and positive parity assignment of the hypertriton ground state were established by the analysis of hypertriton weak decay measurements~\cite{Ke73}. However, the above canonical value of $B^{\rm EMUL}_{\Lambda}(^{3}_{\Lambda}{\rm H})$ has been challenged recently by the STAR Collaboration, claiming much more tightly bound hypertriton with $B^{\rm STAR}_{\Lambda}(^{3}_{\Lambda}{\rm H}) = (0.41\pm 0.12 \pm 0.11)$~MeV~\cite{star20}. 
In Fig.~\ref{interactions}, we show compilation of $\Lambda N$ scattering lengths and $B_\Lambda ({\rm ^3_\Lambda H})$ values obtained in various interaction models. The calculated $B_\Lambda ({\rm ^3_\Lambda H})$ energies are compared with the experimental values $B^{\rm EMUL}_{\Lambda}(^{3}_{\Lambda}{\rm H})$ and $B^{\rm STAR}_{\Lambda}(^{3}_{\Lambda}{\rm H})$ in Fig.~1a.  The figure shows that some former interaction models do not reproduce $B^{\rm EMUL}_{\Lambda}(^{3}_{\Lambda}{\rm H})$, and the NCS97d model even leaves the hypertriton unbound. On the other hand, the scattering lengths $\chi$EFT(NLO)-A,~B,~C in next to leading order chiral effective field theory were tuned to yield the STAR experiment value. It is to be noted that the $\chi$EFT(LO,NLO) $\Lambda$ separation energy in the hypertriton is cutoff dependent (see ref.~\cite{le20}, Table 2) -- in Fig.~1a we present $B_{\Lambda}(^{3}_{\Lambda}{\rm H})$ for cutoff value $\lambda = 500$~MeV. The increased  binding of the hypertriton should affect its lifetime $\tau(^3_{\Lambda}{\rm H})$ and, indeed, the STAR Collaboration reported $\tau(^3_{\Lambda}{\rm H})$  considerably shorter than the free $\Lambda$ lifetime~\cite{star18}.
\\

Since $B_{\Lambda}(^{3}_{\Lambda}{\rm H})$ is used for fine tuning of the $\Lambda$ hypernuclear interactions, obvious questions arise: How does its possibly larger value manifest itself in calculated characteristics of other $\Lambda$  hypernuclei? How does the value of $B_{\Lambda}(^{3}_{\Lambda}{\rm H})$ affect conclusions regarding the recently discussed nature of the $\Lambda NN$ states -- $\Lambda nn$ and $^{3}_{\Lambda}{\rm H}^*(3/2^+)$? The implications of the increased hypertriton binding for next-to-leading order (NLO) $\chi$EFT calculations of A= 4, 5, and 7 $\Lambda$ hypernuclei have been studied by Le et al.~\cite{le20}. While the 3- and 4-body Faddeev and Faddeev-Yakubovski calculations were performed using bare interactions, the 5- and 7-body systems were described within 
the no-core shell model with SRG evolved interactions constrained to reproduce $B_\Lambda({\rm ^5_\Lambda He})$.  These   authors increased the $\Lambda N$ singlet scattering length $a^{\Lambda N}_0$ in order to get the hypertriton binding consistent with the STAR data, while simultaneously reducing the triplet scattering length $a^{\Lambda N}_1$ to preserve the good description of scattering data. This led to 3 sets of the $\Lambda N$ scattering lengths based on the NLO19 interaction model~\cite{hmn20} [we denote them as $\chi$EFT(NLO) - A, B, C]. In view of large uncertainties involved, Le et al.~\cite{le20} have not found any principal reason that would rule out the larger value $B^{\rm STAR}_{\Lambda}(^{3}_{\Lambda}{\rm H})$.

The impact of $B_{\Lambda}(^{3}_{\Lambda}{\rm H})$ on the hypertriton lifetime has been explored recently. Hildenbrand and Hammer~\cite{HH20} found small sensitivity of the hypertriton lifetime to the $\Lambda$ separation energy, but strong  $B_{\Lambda}(^{3}_{\Lambda}{\rm H})$ dependence of the partial widths and the branching ratio $R_3 = \Gamma(^3_{\Lambda}{\rm H} \rightarrow {\rm ^3 He} + \pi^-)/\Gamma_{\pi^-}(^3_{\Lambda}{\rm H})$~\cite{ber70}. 
P\'{e}rez-Obiol et al.~\cite{pgfg20} concluded that 
the hypertriton lifetime $\tau(^3_{\Lambda}{\rm H})$ varied strongly with
$B_{\Lambda}(^{3}_{\Lambda}{\rm H})$ and showed that each of the $\tau(^3_{\Lambda}{\rm H})$ values reported by the  ALICE~\cite{alice19}, HypHI~\cite{rap13} and STAR~\cite{star18} collaborations could be correlated with a theoretically derived value and its own corresponding value of  $B_{\Lambda}(^{3}_{\Lambda}{\rm H})$. Experiments proposed recently at MAMI~\cite{mami18}, JLab~\cite{jlab18}, and J-PARC~\cite{ma} aiming at resolving the 'hypertriton lifetime puzzle' are expected to provide the value of  $B_{\Lambda}(^3_{\Lambda}{\rm H}$) with a resolution better than 50~keV.

In this work we report our study of the consequences of increased $\Lambda$ separation energy in the hypertriton, announced by the STAR collaboration, for selected $s$-shell hypernuclear systems, namely $\Lambda nn (1/2^+)$, $^{3}_{\Lambda}{\rm H}^*(3/2^+)$, and  $^{5}_{\Lambda}$He. 
Since the pioneering calculation~\cite{DHT72}, which revealed overbinding of $^5_{\Lambda}$He while reproducing binding energies of the rest of the $s$-shell hypernuclei, numerous other works have also failed to describe simultaneously the  
few-body hypernuclear bound states. Only recently, Contessi et al.~\cite{CBG18} succeeded to    
solve this "overbinding problem" within a pionless effective field theory at leading order (LO \nopieft). 
The neutral $\Lambda nn$ system has first been found unbound by Downs and Dalitz~\cite{DD59} and since then various few-body calculations supported their conclusion~\cite{garcilazo87, MKGS95, GFV07a, GFV07b, BRS08, GG14, HOGR14, GG19, ARO15, HH19, AG15, SBBM20, SBBM21}. Moreover, most of these works predicted the excited state of the hypertriton, $^3_{\Lambda}{\rm H}^*(3/2^+)$ to be located above the $\Lambda d$ threshold~\cite{MKGS95, GFV07, GFV07b, SBBM20, SBBM21}.  
The recent interest of experimentalists in the nature of the $\Lambda nn$~\cite{JlabE17003} as well as $^3_{\Lambda}{\rm H}^*(3/2^+)$~\cite{JlabP19002} states, i.e., whether they are bound or in continuum, was motivated by the evidence for the bound $\Lambda nn$ state reported by the HypHI collaboration~\cite{HypHI13}. In any case, the study of hypernuclear $\Lambda NN$ trios  provides much-needed information on the spin and isospin dependence of the $\Lambda N$ and $\Lambda NN$ interactions. 

In our present calculations, the two- and three-body interactions among baryons are  described within LO \nopieft. This approach has already been successfully applied to calculations of $s$-shell single-$\Lambda$~\cite{CBG18} and double-$\Lambda$ hypernuclear systems~\cite{CSBGM19}. 
Recent LO \nopieft calculations of hypernuclear trios~\cite{SBBM20, SBBM21} have predicted $\Lambda nn$ and $^{3}_{\Lambda}{\rm H}^*(3/2^+)$ to be unbound,  confirming thus conclusions of previous theoretical analyses. The investigation of hypernuclear  continuum states using the complex scaling method~\cite{AC71} and the inverse analytic continuation in the coupling constant (IACCC) method~\cite{HP17} led to conclusion that $\Lambda nn$ 
exists likely as a subthreshold resonant state while the excited state of the hypertriton, $^3_{\Lambda}{\rm H}^*$, 
is a virtual state located just above the $\Lambda d$ threshold. It is thus quite legitimate to ask whether the 
increased binding of the hypertriton, as claimed by the STAR collaboration, could lead to a bound excited state $^3_{\Lambda}{\rm H}^*$. Similarly, the stronger hypertriton binding could affect the position of the $\Lambda nn$ pole in the complex energy plane and convert it to a true resonance with Re$(E) > 0$.  
The near-threshold virtual state of $^3_{\Lambda}{\rm H}^*$ was found to have a
strong effect on the $\Lambda d$ $s$-wave phase shifts in the $J^{\pi} = 3/2^+$ channel~\cite{SBBM20}. The $\Lambda d$ scattering at energies close to the threshold has been subject of only few theoretical works due to nonexistence of data on $\Lambda$ scattering off deuteron (see~\cite{JH20} and references therein). However, 
the missing information about interaction dynamics from scattering experiments involving hadrons with non-zero strangeness could be provided by measuring two-particle correlation functions in high-energetic $pp$  and heavy-ion collisions~\cite{alice20}. A theoretical study of the $\Lambda d$ momentum correlation functions and their capability to provide additional information on the $\Lambda N$ interaction has been performed recently by Haidenbauer~\cite{JH20}. We address here the issue of $\Lambda d$ correlations as well, concentrating on implications of the increased hypertriton binding.

At LO \nopieft the parameters of the two-body $\Lambda N$ interactions are fixed by the $\Lambda N$ spin-singlet $a_0^{\Lambda N}$ and spin-triplet $a_1^{\Lambda N}$ scattering lengths. Parameters of the three-body $\Lambda NN$ force are fitted to the $B^{\rm EMUL}_{\Lambda}(^{3}_{\Lambda}{\rm H})$ or $B^{\rm STAR}_{\Lambda}(^{3}_{\Lambda}{\rm H})$ value, and the experimental $\Lambda$ separation energies in the 4-body systems $^{4}_{\Lambda}{\rm H}(0^+)$ \cite{4LH0+exp} and $^{4}_{\Lambda}{\rm H}^*(1^+)$ \cite{4LH1+exp}. 
Calculations of hypernuclear bound states are performed within the stochastic variational method (SVM)~\cite{VS95, SV98}, location of the poles of continuum states are determined by the IACCC method. 

In this work, the Coulomb interaction is not included. We use a charge symmetric version of the LO \nopieft hypernuclear interaction for convenience, parameters of the three-body $\Lambda NN$ force in the I=1 and I=0, S=3/2 channels are constrained by $B_\Lambda$ of $\rm ^4_\Lambda H$ and $\rm ^4_\Lambda H^*$. Alternatively, one can consider mean experimental $B_\Lambda$ energies for $\rm ^4_\Lambda H$ and $\rm ^4_\Lambda He$. However, such a change would induce only a minor difference in four-body $\Lambda$ sepration energies of order of the corresponding experimental errors ($\approx 100$~keV). The quantitative effect of charge symmetry breaking in 4-body systems on our results can then be deduced from plots presented below, where it is roughly comparable to uncertainties induced by experimental errors in 4-body $B_\Lambda$ input values.       

The paper is organized as follows: In Section II, we present the model and methodology applied in our calculations of few-body hypernuclear systems. 
The \nopieft approach, as well as the SVM and IACCC methods are described only briefly since they were discussed in sufficient detail in our previous papers~\cite{SBBM20,SBBM21}.   
In Section III, we present results of our investigation of the consequences of the increased $^3_{\Lambda}$H binding. We first demonstrate that in the case of $^5_{\Lambda}$He, the STAR experiment value $B_\Lambda^{\rm STAR}({\rm ^3_\Lambda H})$  is acceptable within the LO truncation error and could lead to $B_\Lambda({\rm ^5_\Lambda He})$ in agreement with experiment. Then we explore the position of the 
$\Lambda nn$ pole in the complex energy plane and find the $\Lambda nn$ resonance to be less likely observed experimentally with increasing $B_{\Lambda}(^3_{\Lambda}$H) as it moves to the unphysical region.  We further discuss how the $\Lambda$ separation energy in $^3_{\Lambda}{\rm H}(1/2^+)$ affects the nature of the hypertriton excited state $^3_{\Lambda}{\rm H}(3/2^+)$ and its position with respect to the $\Lambda d$ threshold.   We show that the near threshold excited (virtual) state could strongly affect the $\Lambda d$ elastic scattering cross-section. Moreover, we observe that the larger value of  
 $B^{\rm STAR}_{\Lambda}(^3_{\Lambda}$H) leads to a larger spin-averaged  $s$-wave $\Lambda d$ correlation function. However, in view of large experimental errors and approximations involved at the moment, it is not possible  to discriminate between   $B_\Lambda^{\rm EMUL}({\rm _\Lambda^3 H})$ and $B_\Lambda^{\rm STAR}({\rm _\Lambda^3 H})$. 
For some $\Lambda N$ scattering lengths, we get a weakly bound hypertriton excited state  $^3_{\Lambda}{\rm H}^*(3/2^+)$ which is subject to weak decay as well as electromagnetic M1 dipole transition to the $1/2^+$ hypertriton ground state. We discuss how the energy of such  hypothetically bound excited state and the corresponding energy splitting between the $\rm ^3_\Lambda H^*(3/2^+)$ and   $\rm ^3_\Lambda H(1/2^+)$ affect its lifetime. 
Finally, we summarize our findings in Section IV.

\section{Model and Methodology}
In this work we describe nuclear and hypernuclear interaction using \nopieft at LO \cite{CBG18,SBBM20,SBBM21}. Within this low-energy approach we consider only nucleon $N$ and $\Lambda$ hyperon degrees of freedom while pions are integrated out. At LO there are four 2-body and four 3-body $s$-wave momentum-independent contact terms, each associated with different $NN$, $\Lambda N$, $NNN$, and $\Lambda NN$ isospin-spin $(I,S)$ channels. The contact terms are then regularized by applying a Gaussian regulator with momentum cutoff $\lambda$. This procedure yields the two-body $V_2$ and three-body $V_3$ parts of the LO \nopieft potential 
\begin{eqnarray}
&V_2 =\sum_{I,S} C_\lambda^{I,S} \sum_{i<j}\mathcal{P}^{I,S}_{ij} {\rm e}^{-\frac{\lambda^2}{4}r_{ij}^2},\nonumber\\ 
&V_3 =\sum_{I,S} D_\lambda^{I,S} \sum_{i<j<k}\mathcal{Q}^{I,S}_{ijk}\sum_{\rm cyc}{\rm e}^{-\frac{\lambda^2}{4}\left(r_{ij}^2+r_{jk}^2\right)}, \label{potential}\\
\nonumber
\end{eqnarray}
where $\mathcal{P}^{I,S}_{ij}$ and $\mathcal{Q}^{I,S}_{ijk}$ are projection operators into individual $s$-wave $(I,S)$ two-body and three-body channels, respectively. Low energy constants (LECs) $C_\lambda^{I,S}$ and $D_\lambda^{I,S}$ are constrained for each cutoff $\lambda$ by low-energy data. Possible \nopieft predictions then exhibit residual $\lambda$ dependence which is suppressed with cutoff approaching the contact limit $\lambda \rightarrow \infty$ \cite{CBG18}.

We note that the two-body $\Lambda N$ LECs account for contributions from both direct $\Lambda N - \Lambda N$ interactions and indirect $\Lambda N - \Sigma N - \Lambda N$ interactions. This is accomplished as detailed below by fitting these LECs to experimentally constrained values of $\Lambda N $ singlet and triplet scattering lengths. Furthermore, $\Lambda N \leftrightarrow \Sigma N$ conversion, known to have significant effect in other approaches to $s$-shell hypernuclear systems \cite{nogga13,HOGR14}, is partially included in the three-body $\Lambda N N$ LECs which are fixed by using $B^{\rm exp}_{\Lambda}$ values in three-body and four-body hypernuclear systems considered in the present approach \cite{CBG18,SBBM21}. We recall that three-body $NNN$ and $\Lambda NN$ force components enter necessarily at LO in \nopieft in order to avert Thomas collapse \cite{thomas35} of few-body baryonic systems. This contrasts with chiral EFT where three-body forces are expected to enter at $\rm N^2LO$ and their effect is assumed to be small in $\rm ^3_\Lambda H$ but sizeable in heavier hypernuclear systems \cite{le20,hmn20}. In order to clarify further this issue, we present and discuss in the Appendix two- and three-body force contributions in $\rm ^3_\Lambda H$, $\rm ^4_\Lambda H$, $\rm ^4_\Lambda H^*$, $\rm ^5_\Lambda He$, evaluated within LO \nopieft for a selected $a^{\Lambda N}$ set and two cutoff values $\lambda$.

Our choice of LECs proceeds as follows. The nuclear LECs $C_\lambda^{I=0,S=1}$, $C_\lambda^{I=1,S=0}$, and $D_\lambda^{I=1/2,S=1/2}$ are fitted to reproduce the deuteron binding energy $B({\rm ^2H}) = 2.2245$~MeV \cite{LA82}, the $NN$ spin-singlet scattering length $a^{NN}_0 =-18.63$~fm \cite{NSC97}, and the triton binding energy $B({\rm ^3H}) = 8.482$~MeV. Hypernuclear LECs $C_\lambda^{I=1/2,S=1}$, $C_\lambda^{I=1/2,S=0}$, $D_\lambda^{I=0,S=1/2}$, $D_\lambda^{I=1,S=1/2}$, and $D_\lambda^{I=0,S=3/2}$ are constrained to reproduce a specific set of the $\Lambda N$ spin-singlet $a^{\Lambda N}_0$ and spin-triplet $a^{\Lambda N}_1$ scattering lengths, the experimental $\Lambda$ separation energies $B^{\rm EMUL}_{\Lambda}({\rm ^3_\Lambda H};1/2^+)$ or $B^{\rm STAR}_{\Lambda}({\rm ^3_\Lambda H};1/2^+)$, $B_\Lambda ({\rm ^4_\Lambda H};0^+) = 2.16(8)$~MeV \cite{4LH0+exp} and the excitation energy $E_{\rm exc} ({\rm ^4_\Lambda H^*};1^+) = 1.09(2)$~MeV \cite{4LH1+exp}. The lack of $\Lambda N$ scattering data does not allow to sufficiently constrain sizes of $a^{\Lambda N}$s, which leads to rather large uncertainty: $a^{\Lambda N}_0 \in \left(-9.0;0 \right)$~fm, $a^{\Lambda N}_1 \in \left(-3.2;-0.8 \right)$ fm (Sechi-Zorn) \cite{SKTB68} or $a^{\Lambda N}_0 = -1.8^{+2.3}_{-4.2}$~fm, $a^{\Lambda N}_1 = -1.6^{+1.1}_{-0.8}$~fm (Alexander et al.) \cite{alexander68}. Consequently, we either use representative $a^{\Lambda N}$ values given by different $YN$ interaction models - NSC97f \cite{NSC97}, $\chi$EFT(LO) \cite{lo06}, $\chi$EFT(NLO) \cite{nlo13}, and $\chi$EFT(NLO) - A,B,C \cite{le20} (see Fig.~1) or we consider $a^{\Lambda N}_0$ and $a^{\Lambda N}_1$ as parameters which are varied within experimentally acceptable range. The advantage of our approach is that one can fix both $\Lambda N$ scattering lengths and the hypertriton ground state energy and thus study in a larger scope the effect of the increased $B_\Lambda({\rm ^3_\Lambda H;1/2^+})$ on predicted properties of the remaining $s$-shell systems $\Lambda nn$, $\rm ^3_\Lambda H^*$, and $\rm ^5_\Lambda He$.

Hypernuclear bound states are described within the Stochastic Variational Method (SVM) \cite{SV98}. Here, the A-body wave fucntion $\Psi$ is expanded in a correlated Gaussian basis \cite{VS95}
\begin{equation} \label{corrgauss}
   \Psi=\sum_i c_i~\psi_i 
       = \sum_i c_i~\hat{\mathcal{A}}
         \left\{{\rm exp}\left(-\frac{1}{2} {\bf x}^T A_i {\bf x}\right) 
                \chi^i_{S M_S} \xi^i_{I M_I} \right\},
\end{equation}
where $\chi^i_{S M_S}$ ($\xi^i_{I M_I}$) denotes the spin (isospin) part, $\hat{\mathcal{A}}$ stands for the antisymmetrization operator over nucleons, and ${\bf x}=({\bf x}_1, ..., {\bf x}_{A-1})$ is a set of Jacobi vectors. 
Each positive-definite symmetric matrix $A_i$ contains $A(A-1)/2$ stochastically selected parameters. Variational coefficients $c_i$ and the corresponding bound state energies are calculated by diagonalizing the Hamiltonian matrix, i.e. solving the generalized eigenvalue problem.

While bound state solutions can be straightforwardly obtained by employing a basis of square-integrable functions, few-body continuum states (virtual states, resonances) posses different asymptotic behavior and further techniques have to be involved. Following our previous work \cite{SBBM20,SBBM21}, we apply the Inverse Analytic Continuation in the Coupling Constant (IACCC) method \cite{HP17} which provides accurate predictions of resonant and virtual state energies in agreement with the Complex Scaling Method \cite{SBBM21}.

In this approach, positions of the $\Lambda nn$ and $\rm ^3_\Lambda H^*$ continuum states are calculated by supplementing the LO \nopieft potential (\ref{potential}) with an auxiliary attractive 3-body potential
\begin{equation}
 V^{\rm IACCC}_3 = d_\alpha^{I,S} \sum_{i<j<k}\mathcal{Q}^{I,S}_{ijk}\sum_{\rm cyc}{\rm e}^{-\frac{\alpha^2}{4}\left(r_{ij}^2+r_{jk}^2\right)},
 \label{v3iaccc}
\end{equation}
which, by virtue of the projection operator $\mathcal{Q}^{I,S}_{ijk}$, affects only specific $(I,S)$ three-body channels - $(1,\frac{1}{2})$ for $\Lambda nn$ and $(0,\frac{3}{2})$ for $\rm ^3_\Lambda H^*$. Here, the value of the range parameter $\alpha$ is always selected to be equal to the value of the \nopieft cutoff $\lambda$. 

Increasing the $V^{\rm IACCC}_3$ attraction, a continuum $S$-matrix pole given by the LO \nopieft potential (\ref{potential}) starts to move towards the lowest threshold and at specific $d_{0,~\alpha}^{I,S}$ value it turns into a bound state. Applying the SVM we calculate a set of $M+N+1$ bound state energies $\{E_{\rm B}^i(d_{i,~\alpha}^{I,S});~d_{i,~\alpha}^{I,S}<d_{0,~\alpha}^{I,S};~i=1,\dots,M+N+1\}$ which are used to construct a Pad\'{e} approximant of degree ($M$,$N$) $\mathcal{P}^{(M,N)}$ of function $d(\kappa)$
\begin{equation}
    \mathcal{P}^{(M,N)}(\kappa) =\frac{\sum_{j=0}^M b_j \kappa^j}{1+\sum_{j=1}^N c_j \kappa^j} \approx d(\kappa).
    \label{padeapprox}
\end{equation}
Here, $\kappa=-{\rm i}k=-{\rm i}\sqrt{E_{\rm B}}$ with $E_{\rm B}$ the bound state energy value, measured with respect to the lowest dissociation threshold. The $b_j$, $c_j$ are real parameters of the $\mathcal{P}^{(M,N)}$. The position of a resonance or virtual state is then determined by setting $d=0$ in Eq.~(\ref{padeapprox}) and by searching for a physical root of the polynomial equation
\begin{equation}
    \sum_{j=0}^M b_j \kappa^j=0.
    \label{poleq}
\end{equation}
The energy of the pole $E=({\rm i}\kappa)^2$ is in general complex. Consequently, for the $\Lambda nn$ resonance we keep the common notation $E=E_r~-~ {\rm i}\Gamma/2$, where $E_r={\rm Re}(E)$ is the position of the resonance and $\Gamma=-2~{\rm Im}(E)$ stands for the resonance width. For the $\rm ^3_\Lambda H^*$ virtual state we use the relative $\Lambda$-deuteron momentum $\gamma_{3/2}=\sqrt{2\mu_{\Lambda d}E}$ in order to avoid confusion between a bound and virtual state energy which are both negative, $\mu_{\Lambda d}=m_\Lambda m_d /(m_\Lambda + m_d)$ stands for the relative mass.   

\section{Results}
The formalism introduced in the previous section was applied in calculations of $s$-shell hypernuclear systems with the aim to explore consequences of increased hypertriton binding for $\Lambda nn$, $\rm ^3_\Lambda H^*$, and $\rm ^5_\Lambda He$. In this section, we present results of the analysis considering both a wide range of $a^{\Lambda N}$ scattering lengths and systematic uncertainties induced by 3- and 4-body experimental constraints. The last part of this section is  dedicated to a hypothetical case of a bound hypertriton excited state $\rm ^3_\Lambda H^*$ located in the vicinity of the hypertriton ground state. Here, we discuss the $\Lambda$ separation energy of $^3_\Lambda {\rm H}^*$ and implications for experimental measurement of its lifetime $\tau({\rm ^3_\Lambda H}^*)$.      

\subsection{$\rm ^5_\Lambda He$ hypernucleus}
\begin{figure*}
\includegraphics[width=\textwidth]{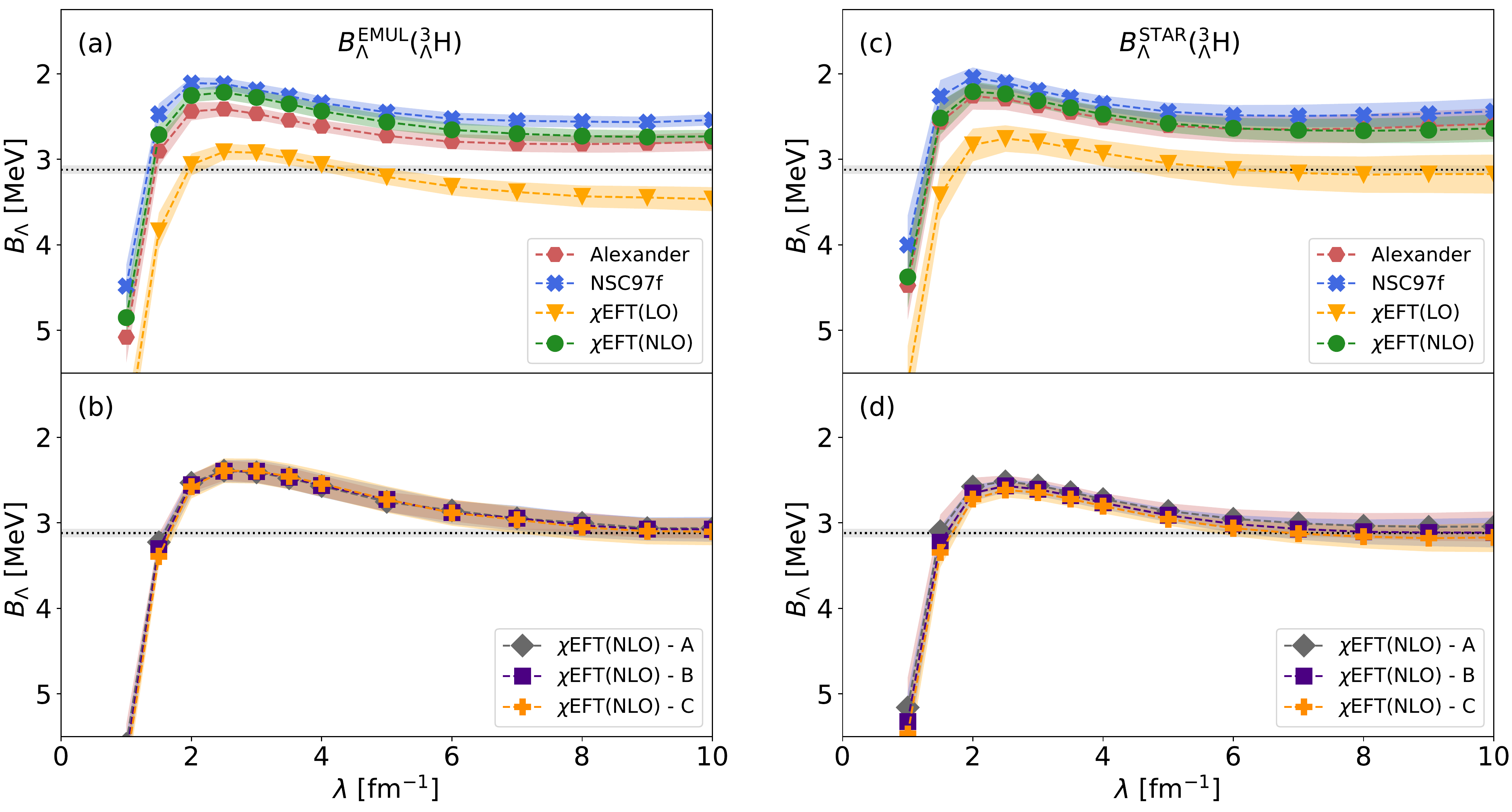}
\caption{$B_\Lambda({\rm ^5_\Lambda He})$ as a function of the cutoff $\lambda$, calculated for several different $\Lambda N$ interaction strengths. The three-body $\Lambda NN$ LEC $D^{I=0,S=1/2}_\lambda$ is fixed to $B^{\rm EMUL}_\Lambda({\rm ^3_\Lambda H})$ (a,b) or $B^{\rm STAR}_\Lambda({\rm ^3_\Lambda H})$ (c,d). The shaded areas denote uncertainty of our calculations induced by experimental errors in 3- and 4-body constraints. Dotted black line marks the experimental value $B^{\rm exp}_\Lambda({\rm ^5_\Lambda He})=3.12(2)$~MeV \cite{davis05} with the corresponding experimental error (narrow shaded area alongside the dotted line).}
\label{5LHe:1}
\end{figure*}
We start our discussion by comparing the LO \nopieft prediction of the $\Lambda$ separation energy in $\rm ^5_\Lambda He$ with the experimental value $B^{\rm exp}_\Lambda({\rm ^5_\Lambda He})=3.12(2)$~MeV \cite{davis05}. In Fig.~\ref{5LHe:1} we present calculated $B_\Lambda({\rm ^5_\Lambda He})$ as a function of the momentum cutoff $\lambda$ for several $a^{\Lambda N}$ sets and two different $B_\Lambda({\rm ^3_\Lambda H})$ constraints - $0.13(5)$~MeV (Figs.~\ref{5LHe:1}a and \ref{5LHe:1}b) and $0.41(12)$~MeV (Figs.~\ref{5LHe:1}c and \ref{5LHe:1}d). Shaded areas indicate systematic uncertainty induced by experimental errors in the 3- and 4-body constraints $B_\Lambda({\rm _\Lambda^3 H})$, $B_\Lambda({\rm _\Lambda^4 H; 0^+})$, and $B_\Lambda({\rm _\Lambda^4 H;1^+})$. We find that despite rather different strengths of the $\Lambda N$ spin-singlet and spin-triplet channels considered, in particular between the Alexander, NSC97f, $\chi$EFT(LO), $\chi$EFT(NLO) sets (Figs.~\ref{5LHe:1}a and \ref{5LHe:1}b) and the $\chi$EFT(NLO)~-~A,~B,~C sets (Figs.~\ref{5LHe:1}b and \ref{5LHe:1}d), calculated $B_\Lambda({\rm ^5_\Lambda He})$ does not change dramatically. We further observe that increasing the $B^{\rm EMUL}_\Lambda({\rm ^3_\Lambda H})$ constraint by $\approx 0.28$~MeV to $B^{\rm STAR}_\Lambda({\rm ^3_\Lambda H})$ has only a minor effect on the predicted $\Lambda$ separation energy 
$B_\Lambda({\rm ^5_\Lambda He})$ (compare the left and right panels of Fig.~\ref{5LHe:1}).
\begin{figure}[b!]
\includegraphics[width=0.5\textwidth]{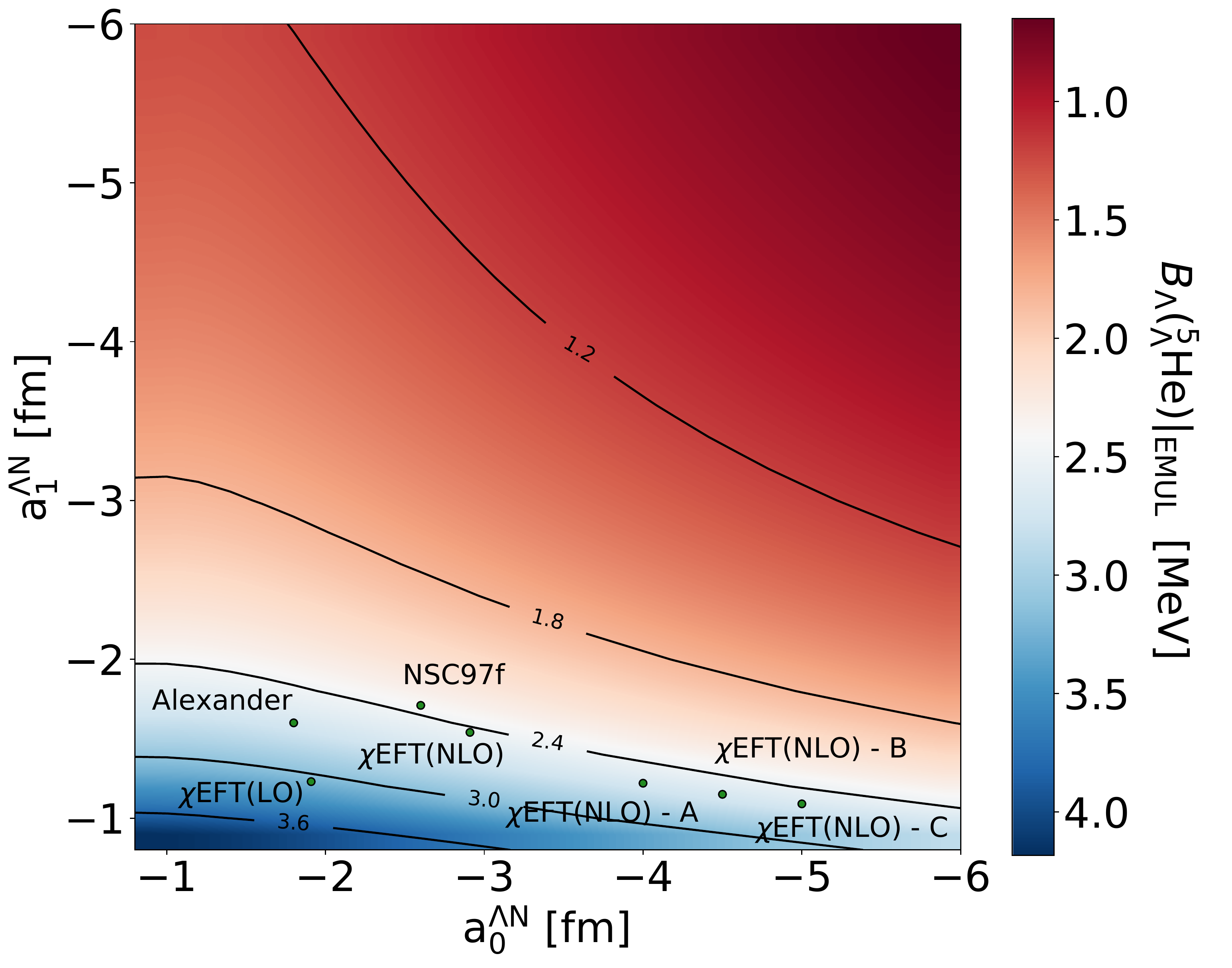}
\caption{$B_\Lambda({\rm ^5_\Lambda He})$ calculated for $\lambda=4~{\rm fm^{-1}}$, the $B^{\rm EMUL}_\Lambda({\rm ^3_\Lambda H})$ three-body constraint, and a wide range of the $a^{\Lambda N}_0$ and $a^{\Lambda N}_1$ scattering lengths. Black dots denote sets of $a^{\Lambda N}$s considered throughout this work.}
\label{5LHe:2}
\end{figure}

\begin{figure*}[t!]
\includegraphics[width=\textwidth]{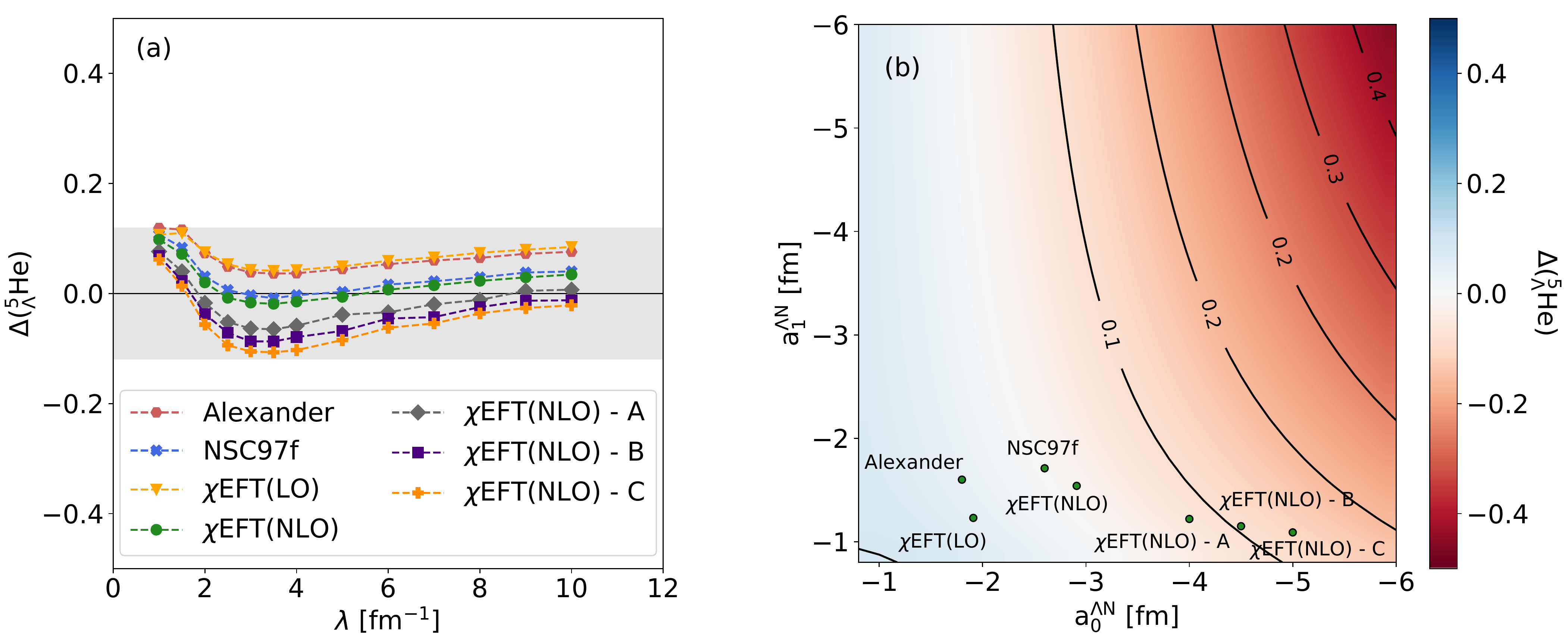}
\caption{Relative difference $\Delta({\rm ^5_\Lambda He})$ (\ref{difference}) as a function of $\lambda$ for selected sets of $a^{\Lambda N}$s (a) and $\Delta({\rm ^5_\Lambda He})$ calculated for a wide range of $a^{\Lambda N}$s and $\lambda =4~{\rm fm^{-1}}$ (b).}
\label{5LHe:3}
\end{figure*}
The presented low sensitivity of $B_\Lambda({\rm ^5_\Lambda He})$ on different strengths of the $\Lambda N$ two-body channels might be related to the specific choice of the $a^{\Lambda N}_0$ and $a^{\Lambda N}_1$ two-body constraints. In order to clarify this point we show in Fig.~\ref{5LHe:2} the $\Lambda$ separation energy $B_\Lambda({\rm ^5_\Lambda He})$, calculated for $\lambda = 4~{\rm fm^{-1}}$, the three-body constraint $B_\Lambda^{\rm EMUL}({\rm ^3_\Lambda H})$, and a wide range of the $\Lambda N$ scattering lengths $a^{\Lambda N}_0, a^{\Lambda N}_1 \in \left<-6;-1\right>$~fm. It is apparent that all $a^{\Lambda N}$ sets considered in Fig.~\ref{5LHe:1} are located within an area given by the condition $a^{\Lambda N}_1 =-0.2~a^{\Lambda N}_0 - 2.08(30)~{\rm fm}$. We stress that though this area is completely independent of LO \nopieft, the $a^{\Lambda N}$ sets surprisingly yield roughly the same LO \nopieft prediction of $B_\Lambda({\rm ^5_\Lambda He})$. As demonstrated in Fig.~\ref{5LHe:2},  $B_\Lambda({\rm ^5_\Lambda He})$ calculated within our approach imposes rather strict constraint on the $\Lambda N$ spin-triplet scattering length. Considering either unusually low or large $a^{\Lambda N}_1$ leads to strong deviation from the experimental value $B^{\rm exp}_\Lambda({\rm ^5_\Lambda He})$.

A rather small change in $B_\Lambda({\rm ^5_\Lambda He})$ caused by different hypertriton ground state constraints $B_\Lambda({\rm ^3_\Lambda H})$ deserves further discussion. In Fig.~\ref{5LHe:3}a we show the relative difference $\Delta({\rm ^5_\Lambda He})$ between $B_\Lambda({\rm ^5_\Lambda He})$ calculated using the $B^{\rm EMUL}_\Lambda({\rm ^3_\Lambda H})$ and $B^{\rm STAR}_\Lambda({\rm ^3_\Lambda H})$ constraints  
\begin{equation}
    \Delta({\rm ^5_\Lambda He})=\frac{B_\Lambda({\rm ^5_\Lambda He})|_{\rm EMUL}-B_\Lambda({\rm ^5_\Lambda He})|_{\rm STAR}}{B_\Lambda({\rm ^5_\Lambda He})|_{\rm EMUL}}.
    \label{difference}
\end{equation}
For all considered $\Lambda N$ potential models and cutoff values, the difference $\Delta({\rm ^5_\Lambda He})$ is less than $\approx 16\%$ (grey shaded area in Fig.~\ref{5LHe:3}a) and evolves only mildly with $\lambda$. 
In~(Fig.~\ref{5LHe:3}b) we present $\Delta({\rm ^5_\Lambda He})$ calculated for $\lambda=4$~fm$^{-1}$ and a wide range of $\Lambda N$ scattering lengths.  
The figure demonstrates only moderate dependence of  $\Delta({\rm ^5_\Lambda He})$ on $a^{\Lambda N}$'s.  The relative difference stays below $16\%$ for all considered $\Lambda N$ interaction models. 
It is to be noted that the variations of the $\Lambda N$ and ${\rm _\Lambda^3 H}$ constraints in LO \nopieft are compensated to some extent by the 4-body constraints  $B_\Lambda ({\rm ^4_\Lambda H};0^+)$ and  $E_{\rm exc} ({\rm ^4_\Lambda H^*};1^+)$. As a results, $B_\Lambda({\rm ^5_\Lambda He})$ is rather insensitive to the increased hypertriton binding energy. This has been demonstrated by changing the $B_\Lambda({\rm _\Lambda^3 H})$ three-body constraint or using rather different $a^{\Lambda N}$ sets.
We expect that larger $\Delta({\rm ^5_\Lambda He})$ or its stronger $a^{\Lambda N}$ dependence would arise as a consequence of a considerably larger hypertriton ground state energy. 

 In order to state whether our calculations in Fig.~\ref{5LHe:1} exclude or support the increased $B_\Lambda ^{\rm STAR}({\rm ^3_\Lambda H})$, one has to take into account the LO truncation error. At LO we do not consider NLO effective range corrections which are expected to be the first significant corrections. Following our recent work \cite{SBBM21}, the effect of higher order terms can be estimated through residual cutoff dependence. Starting at $\lambda \approx 1.25~{\rm fm^{-1}}$, where effective ranges are roughly reproduced, and approaching the contact limit $\lambda \rightarrow \infty$ one obtains an estimate of the LO uncertainty. Based on our $\rm ^5_\Lambda He$ results presented in Figs.~\ref{5LHe:1},~\ref{5LHe:2},~\ref{5LHe:3} we claim that both values $B_\Lambda^{\rm EMUL}({\rm ^3_\Lambda H})$ and  $B_\Lambda^{\rm STAR}({\rm ^3_\Lambda H})$ are acceptable as a \nopieft three-body constraint  within the LO truncation error. They both  lead to $B_\Lambda({\rm ^5_\Lambda He})$ in agreement with its experimental value.

\subsection{$\Lambda nn$ resonance}
\begin{figure}[h!]
\includegraphics[width=0.45\textwidth]{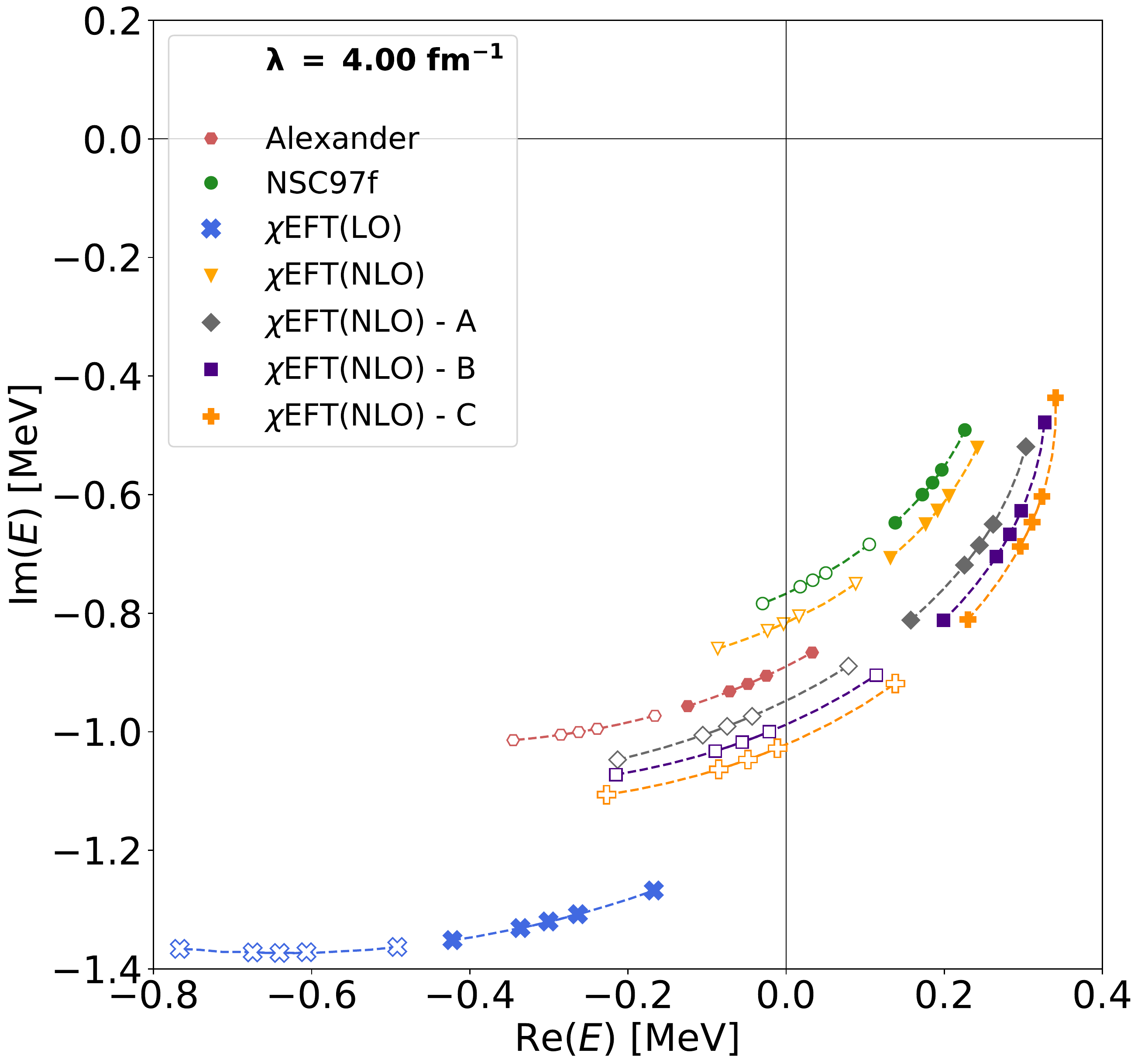}
\caption{Position of the $\Lambda nn$ resonance in the complex energy plane calculated for momentum cutoff $\lambda = 4~{\rm fm^{-1}}$, several $\Lambda N$ interaction strengths, and two different three-body constraints $B_\Lambda^{\rm EMUL}({\rm _\Lambda^3 H})$ (full symbols) and $B_\Lambda^{\rm STAR}({\rm _\Lambda^3 H})$ (empty symbols). The dotted trajectories show uncertainty of calculated $\Lambda nn$ resonance positions induced by experimental errors in few-body constraints - the middle point on each trajectory represents the position with no experimental error considered, two neighbouring points delimit uncertainty caused by errors in 4-body constraints only, while the end points mark the total uncertainty with the experimental error of  $B_\Lambda({\rm _\Lambda^3 H})$ included as well.}
\label{Lnn:1}
\end{figure}

The $\Lambda nn$ resonance position calculated for the momentum cutoff $\lambda =4~{\rm fm^{-1}}$, the $B_\Lambda^{\rm EMUL}({\rm ^3_\Lambda H})$ and $B_\Lambda^{\rm STAR}({\rm ^3_\Lambda H})$ three-body constraints, is presented in Fig.~\ref{Lnn:1}. For all considered $a^{\Lambda N}$ sets, the $\Lambda nn$ resonance position moves with increasing $B_\Lambda({\rm _\Lambda^3 H})$ towards the unphysical third quadrant (Re$(E)<0$, Im$(E)<0$) of the complex energy plane. The physical resonance (Re$(E)>0$, Im$(E)<0$) is convincingly predicted only for $B_\Lambda^{\rm EMUL}({\rm _\Lambda^3 H})$ energy and NSC97f, $\chi$EFT(NLO), and $\chi$EFT(NLO) - A, B, C. For the remaining $a^{\Lambda N}$ interaction strengths, it is located either on the verge between the third and fourth quadrant or deep in the unphysical region.

For each calculation, we studied uncertainty induced by the  experimental errors in 3- and 4-body constraints. The possible $\Lambda nn$ pole position is then represented by the corresponding dashed trajectory. Its shape is exclusively determined by the two-body part of the LO \nopieft Hamiltonian while the specific point on the trajectory is given by the size of the three-body LEC $D_\lambda^{I=1,S=1/2}$ \cite{SBBM20}. There are five different points drawn on each trajectory: The central point represents the calculated $\Lambda nn$ resonance position with no experimental errors taken into account. Two neighbouring points delimit a part of the trajectory given by the uncertainty in 4-body constraints only. The whole trajectory, defined by the last two remaining points, represents the possible $\Lambda nn$ resonance position when all experimental errors in few-body constraints are taken into account. The uncertainty in experimental $B_\Lambda({\rm ^3_\Lambda H})$ has the largest impact on our results, while the influence of experimental errors in the 4-body constraints $B_\Lambda({\rm ^4_\Lambda H;0^+})$ and $E_{\rm exc}({\rm ^4_\Lambda H}; 1^+)$ is minor. 

Our results for $\lambda=4~{\rm fm^{-1}}$ presented in Fig.~\ref{Lnn:1} can be understood as a limiting case which gives the most optimal $\Lambda nn$ resonance position with regard to experiment - a physical $\Lambda nn$ resonance with small width $\Gamma = -2 {\rm Im}(E)$. In Refs.~\cite{SBBM20,SBBM21}, we showed that at LO \nopieft the resonance moves with increasing cutoff from the third unphysical quadrant (Re$(E)<0$, Im$(E)<0$) of the complex energy plane towards the fourth physical one (Re$(E)>0$, Im$(E)<0$) and further closer to the $\Lambda +n+n$ threshold. We also demonstrated that for $\lambda \geq 3~{\rm fm^{-1}}$ there is almost negligible cutoff dependence. Based on our LO calculations we thus estimate that $\Gamma(\Lambda nn) > 0.8~$MeV. This bound takes into account different $B_\Lambda({\rm _\Lambda^3 H})$, experimental errors in the 3- and 4-body constraints, quite large spread of $\Lambda N$ scattering lengths, and $\lambda$ variation which provides roughly the LO truncation error estimate. Within LO \nopieft we thus find highly unlikely that increasing  $B_\Lambda({\rm _\Lambda^3 H})$ would yield either a bound $\Lambda nn$ system or narrow resonance. On the contrary, 
the chance to observe the $\Lambda nn$ resonance in experiment drops with increasing $B_\Lambda({\rm _\Lambda^3 H})$.

\subsection{Excited state of the hypertriton $\rm ^3_\Lambda H^*$}

\begin{figure*}
\includegraphics[width=\textwidth]{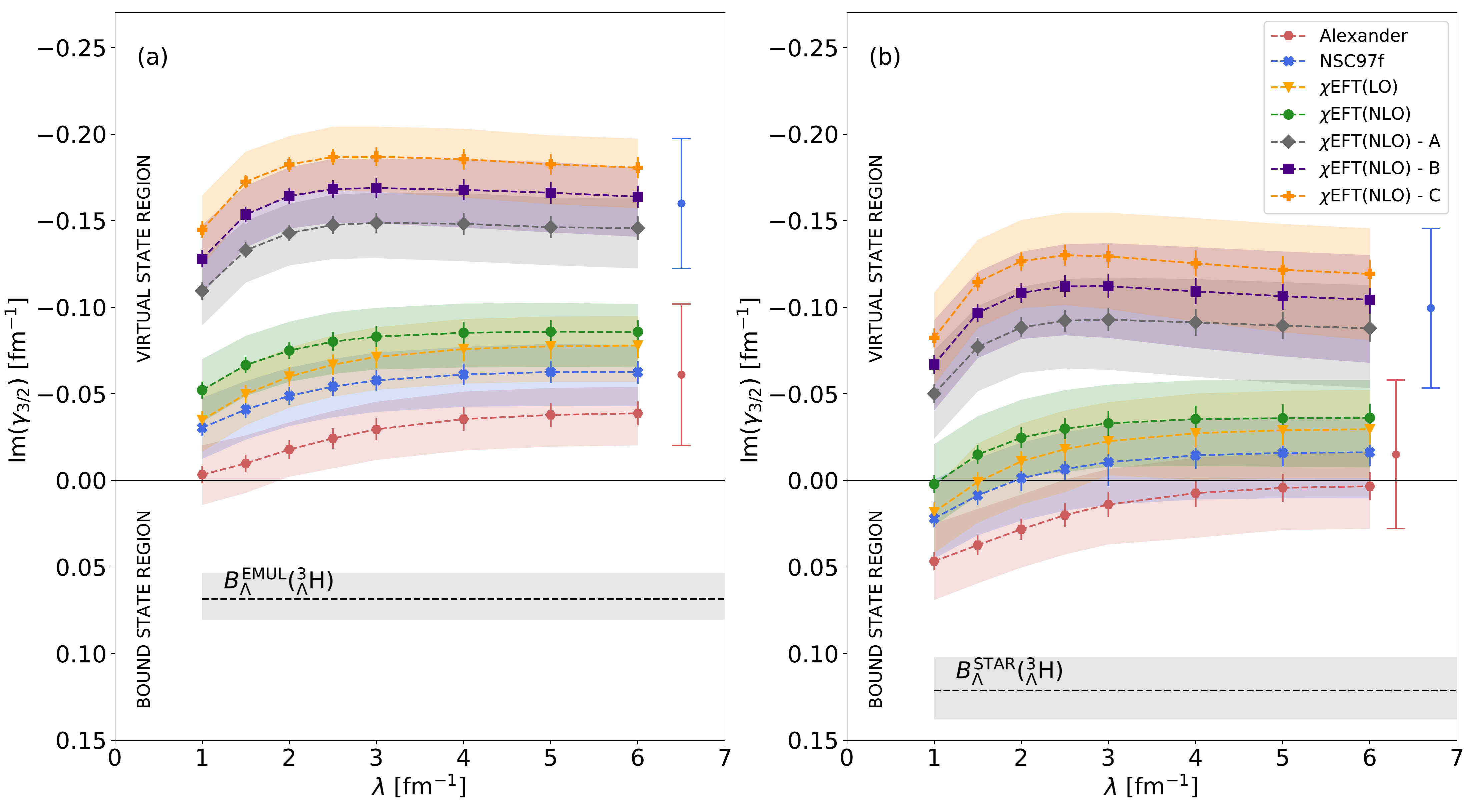}
\caption{Imaginary part of the relative $\Lambda d$ momentum ${\rm Im}(\gamma_{3/2})$ corresponding to a pole of the $\rm ^3_\Lambda H^*$ excited state as a function of cutoff $\lambda$, calculated for several $\Lambda N$ interaction strengths and two values of the $B_\Lambda({\rm ^3_\Lambda H})$ three-body constraint - $B_\Lambda^{\rm EMUL}({\rm _\Lambda^3 H})$ (a) and $B_\Lambda^{\rm STAR}({\rm _\Lambda^3 H})$ (b). Error bars indicate uncertainty of our calculations induced by experimental errors in 4-body constraints. Shaded areas show the total uncertainty induced by both the errors in 4-body constraints and the experimental error in the $B_\Lambda({\rm ^3_\Lambda H})$ three-body constraint. ${\rm Im}(\gamma_{3/2})>0$ represents the bound state region and ${\rm Im}(\gamma_{3/2})<0$ the virtual state region. Dotted lines with shaded areas indicate ${\rm Im}(\gamma_{1/2})=\sqrt{2\mu_{\Lambda d}B_\Lambda({\rm _\Lambda^3 H})}$ binding momentum of the $1/2^+$ hypertriton ground state corresponding to $B_\Lambda^{\rm EMUL}({\rm _\Lambda^3 H})$ (a) and $B_\Lambda^{\rm STAR}({\rm _\Lambda^3 H})$ (b). }
\label{3LH32:1}
\end{figure*}

The $J^\pi=3/2^+$ spin-flip excited state of the hypertriton $\rm ^3_\Lambda H^*$ is the next $s$-shell $\Lambda$-hypernuclear system that can be described within our approach. In Fig.~\ref{3LH32:1} we present the imaginary part of the $\rm ^3_{\Lambda}{\rm H}^*$ pole momentum ${\rm Im}(\gamma_{3/2})$ with respect to the $\Lambda d$ threshold as a function of cutoff $\lambda$, calculated for several $\Lambda N$ interaction strengths and two different hypertriton constraints - $B_\Lambda^{\rm EMUL}({\rm _\Lambda^3 H})$ (Fig.~\ref{3LH32:1}a) and $B_\Lambda^{\rm STAR}({\rm _\Lambda^3 H})$ (Fig.~\ref{3LH32:1}b). It is to be noted that we get ${\rm Re}(\gamma_{3/2})=0$ in all considered cases. 
The $\rm ^3_\Lambda H^*$ is obtained predominantly unbound in a form of a virtual state (${\rm Im}(\gamma_{3/2})<0$). Larger  $B_\Lambda({\rm _\Lambda^3 H})$ shifts the excited state position towards the bound state region (${\rm Im}(\gamma_{3/2})>0$), thus allowing existence of a shallow bound state of $\rm ^3_\Lambda H^*$ for specific $\Lambda N$ interaction strengths   (Fig.~\ref{3LH32:1}b).
\begin{figure*}
\includegraphics[width=0.8\textwidth]{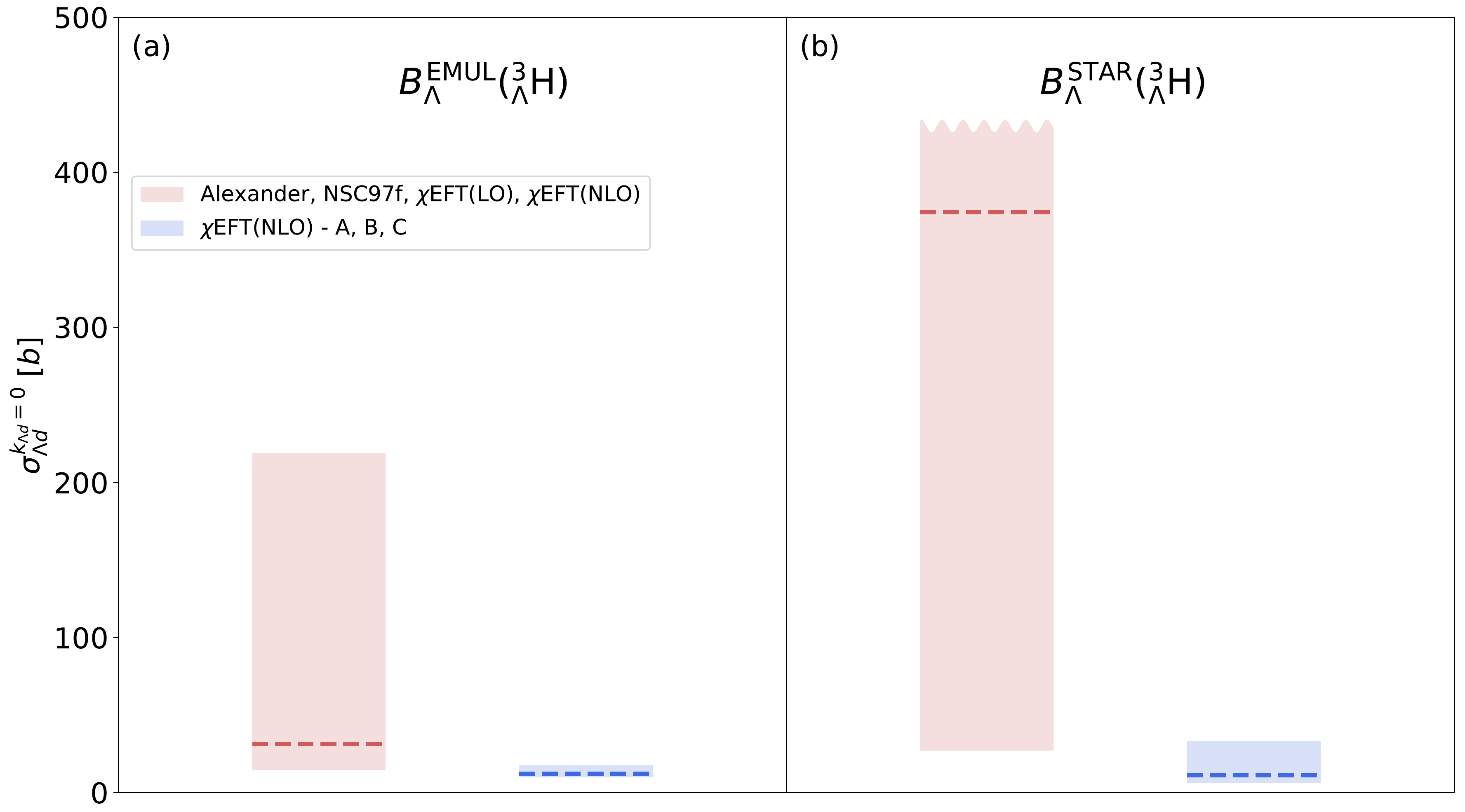}
\caption{Spin-averaged s-wave $\Lambda d$ cross-section at $k_{\Lambda d}=0$ (\ref{crosssection}) calculated using the $B_\Lambda^{\rm EMUL}({\rm _\Lambda^3 H})$ (a) and $B_\Lambda^{\rm STAR}({\rm _\Lambda^3 H})$ (b) three-body constraint.}
\label{3LH32:2}
\end{figure*}

In LO \nopieft the position of the excited state $\rm ^3_\Lambda H^*$ is purely determined by the interplay between the attractive $\Lambda N$ spin-triplet interaction and the repulsive $S=3/2, I=0$ $\Lambda NN$ interaction. We observe that due to few-body dynamics in 3- and 4-body systems, where the corresponding  $D^{I=0,S=3/2}_\lambda$ LEC is fitted, the three-body repulsion becomes stronger with increasing strength of the $\Lambda N$ spin-singlet interaction. Consequently, $a^{\Lambda N}$  sets with a rather large spin-singlet and slightly reduced spin-triplet part, such as $\chi$EFT(NLO)-A,B,C, predict the position of the $\rm ^3_\Lambda H^*$ virtual state farther in continuum.

The uncertainty of the $\gamma_{3/2}$ pole momentum is primarily given by the range of  considered $\Lambda N$ interaction strengths  and  residual cutoff dependence.  Taking into account errors in few-body experimental constraints we observe that the largest contribution  comes from the $B_\Lambda({\rm ^3_\Lambda H})$ experimental error. This can be clearly seen from the comparison between shaded areas [$B_\Lambda({\rm ^3_\Lambda H})$, $B_\Lambda({\rm ^4_\Lambda H; 0^+})$, $B_\Lambda({\rm ^4_\Lambda H; 1^+})$ experimental errors taken into account] and error bars [$B_\Lambda({\rm ^4_\Lambda H; 0^+})$ and $B_\Lambda({\rm ^4_\Lambda H; 1^+})$ errors only].

\begin{figure*}
\includegraphics[width=0.75\textwidth]{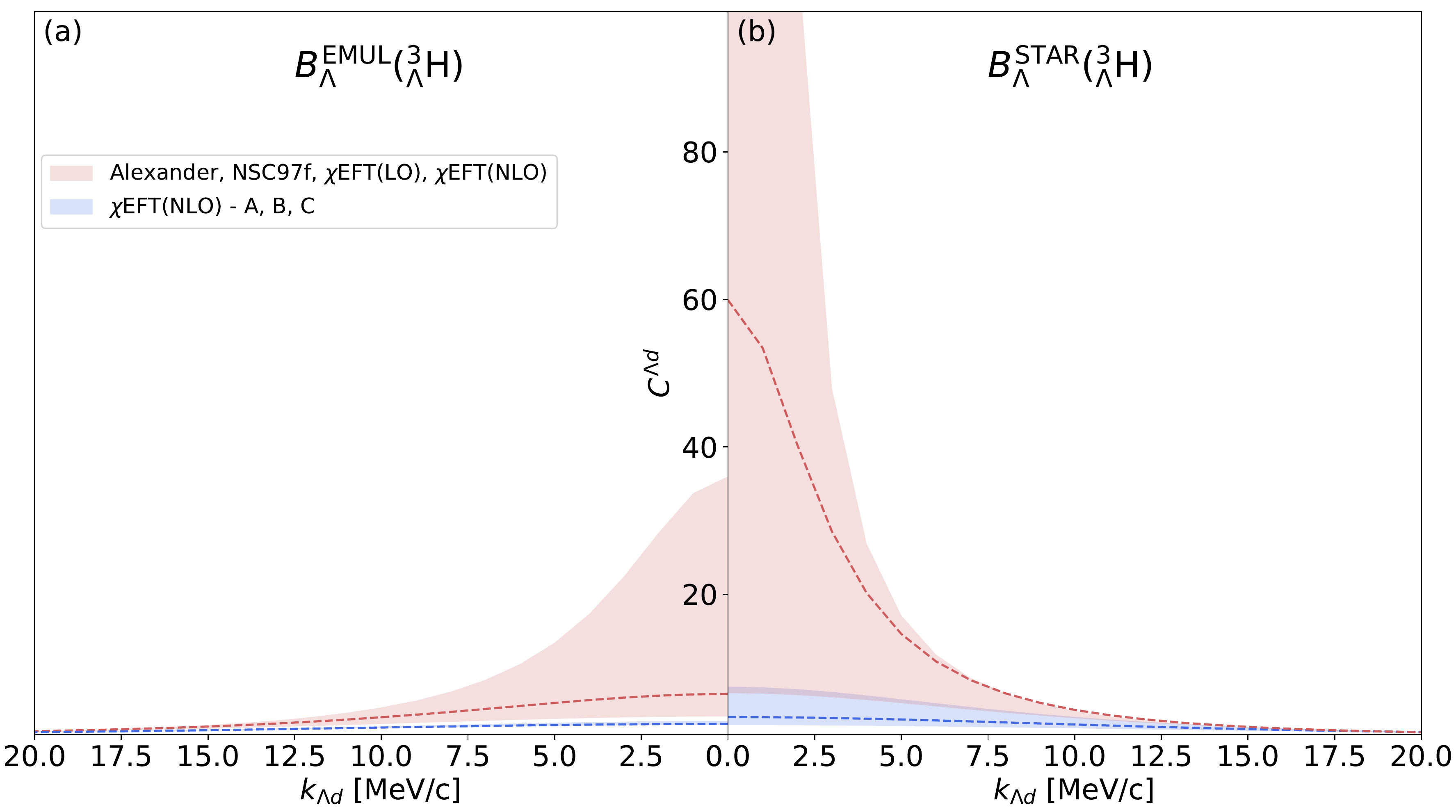}
\caption{Spin-averaged $\Lambda d$ correlations $C^{\Lambda d}$ (\ref{correlations}) for the source size $R=5~{\rm fm}$ as a function of the relative momentum $k_{\Lambda d}$.   }
\label{3LH32:3}
\end{figure*}
Existence of a $\rm ^3_\Lambda H^*$ pole close to the $\Lambda d$ threshold would strongly affect $\Lambda d$ scattering at low energy. Naturally, the most direct probe of its location would involve measurement of the $\Lambda d$ elastic scattering, however, such experiments are highly difficult to conduct and to the best of our knowledge there are none planned in near future. For exploratory reasons, we discuss here the $s$-wave $\Lambda d$ cross-section $\sigma_{\Lambda d}^{k_{\Lambda d}=0}$ at zero relative momentum $k_{\Lambda d}$. For $k_{\Lambda d}=0$ the cross-section can be evaluated via spin-averaged squares of the spin-doublet $A_{\Lambda d}^2(1/2^+)$ and spin-quartet $A_{\Lambda d}^2(3/2^+)$ scattering lengths which can be further estimated using the $\Lambda$ separation energy of the hypertriton ground state and ${\rm Im}(\gamma_{3/2})$ momentum     
\begin{eqnarray}
    \sigma_{\Lambda d}^{k_{\Lambda d}=0}=&4\pi \left[ \frac{1}{3}A_{\Lambda d}^2(1/2^+) + \frac{2}{3} A_{\Lambda d}^2(3/2^+)\right] \nonumber\\
    \simeq & 4\pi \left[\frac{1}{3}\frac{1}{2\mu_{\Lambda d}B_\Lambda({\rm ^3_\Lambda H}; 1/2^+)} +\frac{2}{3}\frac{1}{{\rm Im}(\gamma_{3/2})^2} \right], 
    \label{crosssection}
\end{eqnarray}
    where $\mu_{\Lambda d}$ is the $\Lambda d$ reduced mass.
    While the $\Lambda d$ spin-doublet contribution is fixed either by the $B_\Lambda^{\rm EMUL}({\rm _\Lambda^3 H})$ or $B_\Lambda^{\rm STAR}({\rm _\Lambda^3 H})$ constraint, the spin-quartet contribution comes out as a prediction. As demonstrated in Fig.~\ref{3LH32:1}, the uncertainties in ${\rm Im}(\gamma_{3/2})$, caused by the residual $\lambda$ dependence and experimental errors in few-body constraints, allow us to distinguish only between two groups of the results - for the  Alexander, NSC97f, $\chi$EFT(LO), $\chi$EFT(NLO) sets with roughly comparable $\Lambda N$ spin-singlet and spin-triplet interaction strengths and the $\chi$EFT(NLO) - A, B, C sets where the $\Lambda N$ spin-singlet interaction is enhanced at the expense of the reduced spin-triplet one. The spin-quartet contribution to $\sigma_{\Lambda d}^{k_{\Lambda d}=0}$ is estimated using two groups of the $\rm ^3_\Lambda H^*$ virtual state momenta given by the first and second band of the results at  $\lambda=6~{\rm fm^{-1}}$, depicted aside in Fig.~\ref{3LH32:1} by red and blue error bars, respectively. 
    The distance of the two bands from the $\Lambda d$ threshold is given by the spin-quartet interaction.  
    In Fig.~\ref{3LH32:2} we present $\sigma_{\Lambda d}^{k_{\Lambda d}=0}$ calculated using Eq.~\ref{crosssection} for the two momenta ${\rm Im}(\gamma_{3/2})$ bands and $B_\Lambda^{\rm EMUL}({\rm _\Lambda^3 H})$ (Fig.~\ref{3LH32:2}a), $B_\Lambda^{\rm STAR}({\rm _\Lambda^3 H})$ (Fig.~\ref{3LH32:2}b).  As expected, the spin-quartet contribution and thus the position of the $\rm ^3_\Lambda H^*$ pole with respect to the $\Lambda d$ threshold affects strongly the $\sigma_{\Lambda d}^{k_{\Lambda d}=0}$ cross-section.  
    For the Alexander, NSC97f, $\chi$EFT(LO), $\chi$EFT(NLO) sets of $\Lambda N$ interaction strengths we predict a possibility of exceptionally large $\sigma_{\Lambda d}^{k_{\Lambda d}=0}$ 
    [particularly for $B_\Lambda^{\rm STAR}({\rm _\Lambda^3 H})$], which is not the case for $\chi$EFT(NLO) - A, B, C. Our results thus demonstrate that even limited experimental information about $\sigma_{\Lambda d}$ at low energies, revealing its magnitude, might be efficiently used as an additional constraint to the underlying $\Lambda N$ interaction.

Another possibility how to probe the $\Lambda d$ interaction at low energies is to measure the corresponding correlation function $C^{\Lambda d}(k_{\Lambda d})$. Following the recent work by Haidenbauer \cite{JH20}, we express the $s$-wave correlation function $C^{\Lambda d}(k_{\Lambda d})$ averaged over the spin-doublet $C^{\Lambda d}_{1/2^+}(k_{\Lambda d})$ and spin-quartet $C^{\Lambda d}_{3/2^+}(k_{\Lambda d})$ parts as
\begin{equation}
    C^{\Lambda d}(k_{\Lambda d}) = 1 + \frac{1}{3}C^{\Lambda d}_{1/2^+}(k_{\Lambda d})+\frac{2}{3}C^{\Lambda d}_{3/2^+}(k_{\Lambda d}). \label{correlations}
\end{equation}
Both spin contributions are then evaluated applying the Lednicky-Lyuboshits approach \cite{LL82}, 
\begin{eqnarray}
   &C^{\Lambda d}_{J^\pi}(k_{\Lambda d}) \simeq  \frac{\left|f_{J^\pi}(k_{\Lambda d})\right|^2}{2R^2}F_0(r_{\Lambda d}(J^\pi))\\ 
    &+\frac{2{\rm Re}(f_{J^\pi}(k_{\Lambda d}))}{\sqrt{\pi}R}F_1(2k_{\Lambda d}R)
    -\frac{{\rm Im}(f_{J^\pi}(k_{\Lambda d}))}{R}F_2(2k_{\Lambda d}R) \nonumber
\end{eqnarray}
where $F_0(r_{\Lambda d}(J^\pi))=1-r_{\Lambda d}(J^\pi)/(2\sqrt{\pi}R)$, $F_1(x)=\int_0^x {\rm d}t~e^{t^2-x^2}/x$, and $F_2(x)=(1-e^{-x^2})/x$. The scattering amplitude $f_{J^\pi}(k_{\Lambda d})$ in $C^{\Lambda d}_{J^\pi}(k_{\Lambda d})$ is approximated by the effective range expansion using the $\Lambda d$ scattering lengths $A_{\Lambda d}(1/2^+)$, $A_{\Lambda d}(3/2^+)$ from Eq.~\ref{crosssection} and effective ranges $r_{\Lambda d}(1/2^+)$, $r_{\Lambda d}(3/2^+)$ corresponding to the doublet and quartet channels. The $R$ represents the size of a source approximated by a spherical Gauss function \cite{OMMH16}. In this work we are primarily interested in the low momentum limit where effective range contributions to $C^{\Lambda d}$ are at the level of a small correction. Therefore, we consider fixed values $r_{\Lambda d}(1/2^+)=3~{\rm fm}$ and $r_{\Lambda d}(3/2^+)=4~{\rm fm}$ motivated by the works of Cobis et al. \cite{cobis97} and Hammer \cite{hammer02} (doublet) and our previous work \cite{SBBM20} (spin-quartet channel). For $R$ smaller than the range of an interaction the Lednicky-Lyuboshits approach starts to deviate from the full solution \cite{cho17}, consequently, in order to demonstrate effect of calculated $\rm ^3_\Lambda H^*$ position on $C^{\Lambda d}(k_{\Lambda d})$ we use $R=5~{\rm fm}$.
\begin{figure*}[t!]
\includegraphics[width=\textwidth]{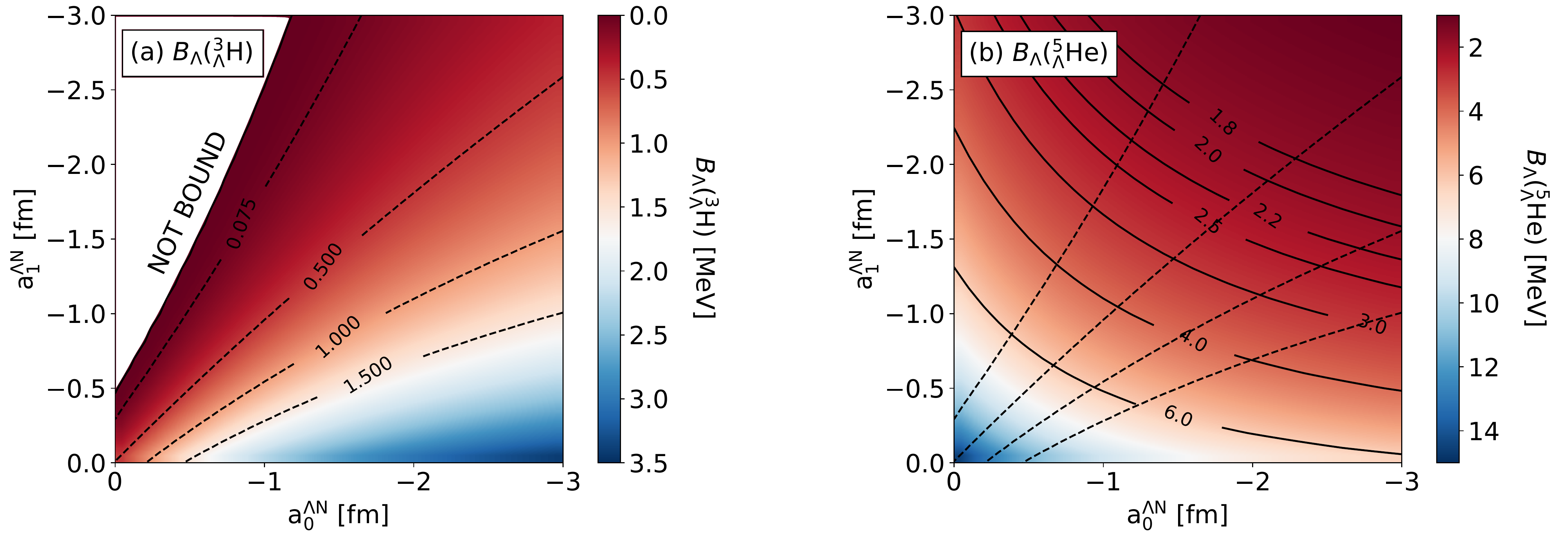}
\caption{The LO \nopieft predictions of hypertriton $1/2^+$ state $B_\Lambda({\rm _\Lambda^3 H})$ (a) and $B_\Lambda({\rm _\Lambda^5 He})$ (b) for wide range of $\Lambda N$ spin-singlet $a^{\Lambda N}_0$ and spin-triplet $a^{\Lambda N}_1$ scattering lengths and $\lambda = 1.5~{\rm fm^{-1}}$. The three-body $D^{I=0,S=1/2}_\lambda$, $D^{I=1,S=1/2}_\lambda$, and $D^{I=0,S=3/2}_\lambda$ LECs are fitted to reproduce simultaneously shallow $3/2^+$ hypertriton bound state with $B_\Lambda({\rm ^3_\Lambda H^*})=0.075~{\rm MeV}$, and experimental $B_\Lambda^{\rm exp}({\rm ^4_\Lambda H}; 0^+)$, $E^{\rm exp}_{\rm exc}({\rm ^4_\Lambda H^*})$. The dashed lines in the left (a) and right (b) panels connect points with the same $B_\Lambda({\rm _\Lambda^3 H})$ while solid lines in the right panel (b) connect points with the same $B_\Lambda({\rm _\Lambda^5 He})$.}
\label{3LH32B:1}
\end{figure*}
\begin{figure*}
\includegraphics[width=\textwidth]{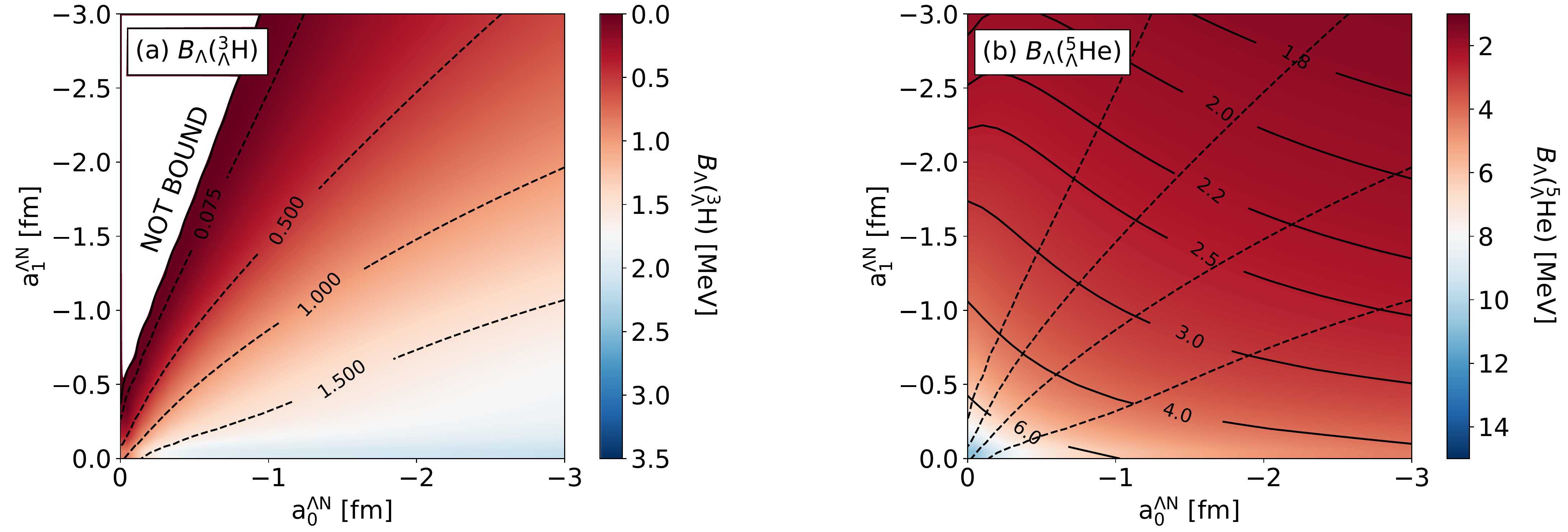}
\caption{The same as Fig.~\ref{3LH32B:1} but for cutoff $\lambda = 4~{\rm fm^{-1}}$.}
\label{3LH32B:2}
\end{figure*}

In Fig.~\ref{3LH32:3} we show the spin-averaged $s$-wave correlation function  $C^{\Lambda d}(k_{\Lambda d})$ for the $B_\Lambda^{\rm EMUL}({\rm _\Lambda^3 H})$ (Fig.~\ref{3LH32:3}a) and $B_\Lambda^{\rm STAR}({\rm _\Lambda^3 H})$ (Fig.~\ref{3LH32:3}b) three-body constraints, calculated using the two groups of the $A_{\Lambda d}(3/2^+)$ values corresponding to (i) the Alexander, NSC97f, $\chi$EFT(LO), $\chi$EFT(NLO)  sets and (ii) the $\chi$EFT(NLO) - A, B, C sets of $\Lambda N$ interaction strengths. Clearly, the magnitude of $C^{\Lambda d}(k_{\Lambda d})$ at low momenta is strongly affected by the distance of both the $1/2^+$ hypertriton ground state and $3/2^+$ excited state from the $\Lambda d$ threshold. We observe that for $B_\Lambda^{\rm STAR}({\rm _\Lambda^3 H})$ the spin-quartet state exists closer or directly passes the threshold, which could lead to large $C^{\Lambda d}(k_{\Lambda d})$ in low momentum region. However, large uncertainties in calculated $C^{\Lambda d}(k_{\Lambda d})$ do not allow to discriminate between the $B_\Lambda^{\rm EMUL}({\rm _\Lambda^3 H})$ and $B_\Lambda^{\rm STAR}({\rm _\Lambda^3 H})$ values. Certainly, a different situation could happen if more precise experimental data on $B_\Lambda({\rm _\Lambda^3 H})$ and $C^{\Lambda d}(k_{\Lambda d})$ are available. This would allow to extract the spin-quartet component which provides constraints on the location of the $\rm ^3_\Lambda H^*$ state. For example, our results connect its  position in continuum further from the $\Lambda+d$ threshold with a considerably larger $\Lambda N$ spin-singlet strength with respect to the spin-triplet channel.

\subsection{Bound excited state of the hypertriton $\rm ^3_\Lambda H^*$ ?}
Constraining the $\Lambda$ separation energy in the $1/2^+$ hypertriton to $B_\Lambda^{\rm STAR}({\rm ^3_\Lambda H})$ (and in very limited cases also to $B_\Lambda^{\rm EMUL}({\rm ^3_\Lambda H})$) leads for specific sets of $a^{\Lambda N}$ scattering lengths to a weakly bound  $3/2^+$ hypertriton excited state (see Fig.~\ref{3LH32:1}). In order to study the hypothetically bound $\rm ^3_\Lambda H^*$ in more detail, we fit our LECs for two cutoff values $\lambda = 1.5~{\rm fm^{-1}}$ and $\lambda = 4.0~{\rm fm^{-1}}$ to $B_\Lambda({\rm ^3_\Lambda H^*})=0.075~{\rm MeV}$,  $B_\Lambda^{\rm exp}({\rm ^4_\Lambda H}; 0^+)$, $E^{\rm exp}_{\rm exc}({\rm ^4_\Lambda H^*})$ and a wide range of $a^{\Lambda N}$ scattering lengths. In this approach, both $B_\Lambda({\rm ^3_\Lambda H})$ and $B_\Lambda({\rm ^5_\Lambda He})$ are predictions of the theory and might be compared to experimental data. In Figs.~\ref{3LH32B:1}~and~\ref{3LH32B:2} we show corresponding $B_\Lambda$ values calculated for different $\Lambda N$ spin-singlet and spin-triplet scattering lengths. Reasonable $B_\Lambda({\rm ^3_\Lambda H})$ energies, in agreement with experiment, are obtained for $|a^{\Lambda N}_1|\gtrsim|a^{\Lambda N}_0|$. Inspecting further  $B_\Lambda({\rm ^5_\Lambda He})$ energy as well, we obtain reasonable predictions for $-1> a^{\Lambda N}_1 > -2~{\rm fm}$, where the acceptable  $a^{\Lambda N}_1$ region slightly changes with cutoff $\lambda$. Rather unusual scenario of stronger $\Lambda N$ spin-triplet interaction is not excluded by scattering data which currently do not provide information on $\Lambda N$ spin-dependence. The result is contradictory to most of $\Lambda N$ interaction models (see Fig.\ref{interactions}) which directly enforce $|a^{\Lambda N}_1|<|a^{\Lambda N}_0|$ to ensure a $1/2^+$ hypertriton ground state. Contrary to these models LO \nopieft includes additional three-body forces, which allows us to find such a solution where $\rm ^3_\Lambda H^*$ is bound and $B_\Lambda$s of the remaining $s$-shell hypernuclei are described reasonably well.

The energy of the hypothetically bound excited state of the hypertriton $\rm ^3_\Lambda H^*$ and the corresponding energy splitting between the $\rm ^3_\Lambda H^*(3/2^+)$ and   $\rm ^3_\Lambda H(1/2^+)$ affect its lifetime. 
The excited state $\rm ^3_\Lambda H^*(3/2^+)$ might  decay both through the weak decay of $\Lambda \rightarrow N\pi$ and through the electromagnetic M1 dipole transition to the $1/2^+$ hypertriton ground state. Following closure-approximation expressions summarized in Ref.~\cite{GG19}, 
weak-decay (WD) rates of $\rm ^3_\Lambda H$ states are given in terms of the 
free-$\Lambda$ WD rate $\Gamma_\Lambda$ by 
\begin{equation} 
\Gamma_{\rm WD}({\frac{1}{2}}^+)/\Gamma_\Lambda=1.114\times 
[|s_{\pi}|^2(1+\frac{1}{2}\eta)+|p_{\pi}|^2(1-\frac{5}{6}\eta)], 
\label{eq:L3Hg.s.} 
\end{equation}
\begin{equation} 
\Gamma_{\rm WD}({\frac{3}{2}}^+)/\Gamma_\Lambda=1.114\times 
[|s_{\pi}|^2(1-\eta)+|p_{\pi}|^2(1-\frac{1}{3}\eta)], 
\label{eq:L3Hexc.} 
\end{equation} 
where the pion decay closure momentum was taken equal to the $\Lambda\to N\pi$ 
decay momentum. The factor 1.114 arises from phase-space factors, $|s_{\pi}|^2
\approx 0.83$ and $|p_{\pi}|^2\approx 0.17$ are parity-violating and 
parity-conserving weights determined in the $\Lambda\to N\pi$ decay, 
and $\eta = 0.13\pm 0.02$ is a strong-interaction exchange integral
ensuring that the summation on final nuclear states is limited to totally 
antisymmetric states (its 0.02 uncertainty reflects the quoted uncertainty 
in the binding-energy value $B_{\Lambda}({\rm ^3_\Lambda H})=0.13\pm0.05$~MeV). 
Here we disregarded pion FSI enhancement, of order 10\% \cite{GG19} to 15\% 
\cite{pgfg20}, expecting most of it to be cancelled out by $\sim$10\% 
interference loss from $\Sigma NN$ small components in the dominantly 
$\Lambda NN$-made $\rm ^3_\Lambda H$ \cite{pgfg20}. Non-pionic decay rate contributions of 
order $\sim$2\% are neglected as well. 

Eqs.(\ref{eq:L3Hg.s.},\ref{eq:L3Hexc.}) for the WD rates, inverse 
of the corresponding WD lifetimes of $\rm ^3_\Lambda H$ and 
$\rm ^3_\Lambda H^*$ states, differ by less than 20\%. For 
a representative value of $\eta =0.13$ we get 
\begin{equation} 
\Gamma_{\rm WD}({\frac{1}{2}}^+)/\Gamma_\Lambda=1.154, \,\,\,\,\, 
\Gamma_{\rm WD}({\frac{3}{2}}^+)/\Gamma_\Lambda=0.986,  
\label{eq:eta} 
\end{equation}
indicating lifetimes close to the free $\Lambda$ lifetime $\tau_{\Lambda} = 263 \pm 2$~ps. 

Assuming the same $1s_{\Lambda}$ w.f., apart from Pauli-spin, for both 
$\rm ^3_\Lambda H^*$ and $\rm ^3_\Lambda H$ doublet levels built on a $^3S_1$ deuteron core, the M1-dominated 
e.m. ${\rm ^3_\Lambda H^*}(3/2^+) \to {\rm ^3_\Lambda H}(1/2^+)$ decay 
rate $\Gamma_{\rm M1}$ is given by~\cite{DG78} 
\begin{equation} 
\Gamma_{\rm M1}=\alpha (\Delta E)^3 \frac{1}{3} (g_c - g_{\Lambda})^2 \, 
{\rm s}^{-1}, 
\label{eq:M1}
\end{equation}
where $\alpha=4.2\cdot 10^{12}\,{\rm s}^{-1}\,{\rm MeV}^{-3}$, $\Delta E$ is 
the deexcitation energy in MeV, and $g_c=0.857$ and $g_{\Lambda}=-1.226\pm 
0.008$ are the gyromagnetic ratios of the 1$^+$ core and the ${\frac{1}{2}}^+$ 
$\Lambda$ hyperon. 

We now combine the separate WD and M1 deexcitation rates to get the overall lifetime 
of $\rm ^3_\Lambda H^*$  
\begin{equation} 
\tau({\frac{3}{2}}^+)=\left (\frac{1}{\tau_{\rm M1}}+\frac{1}{\tau_{\rm WD}}
\right )^{-1}. 
\label{eq:tau1} 
\end{equation}

\begin{figure}
\includegraphics[width=0.45\textwidth]{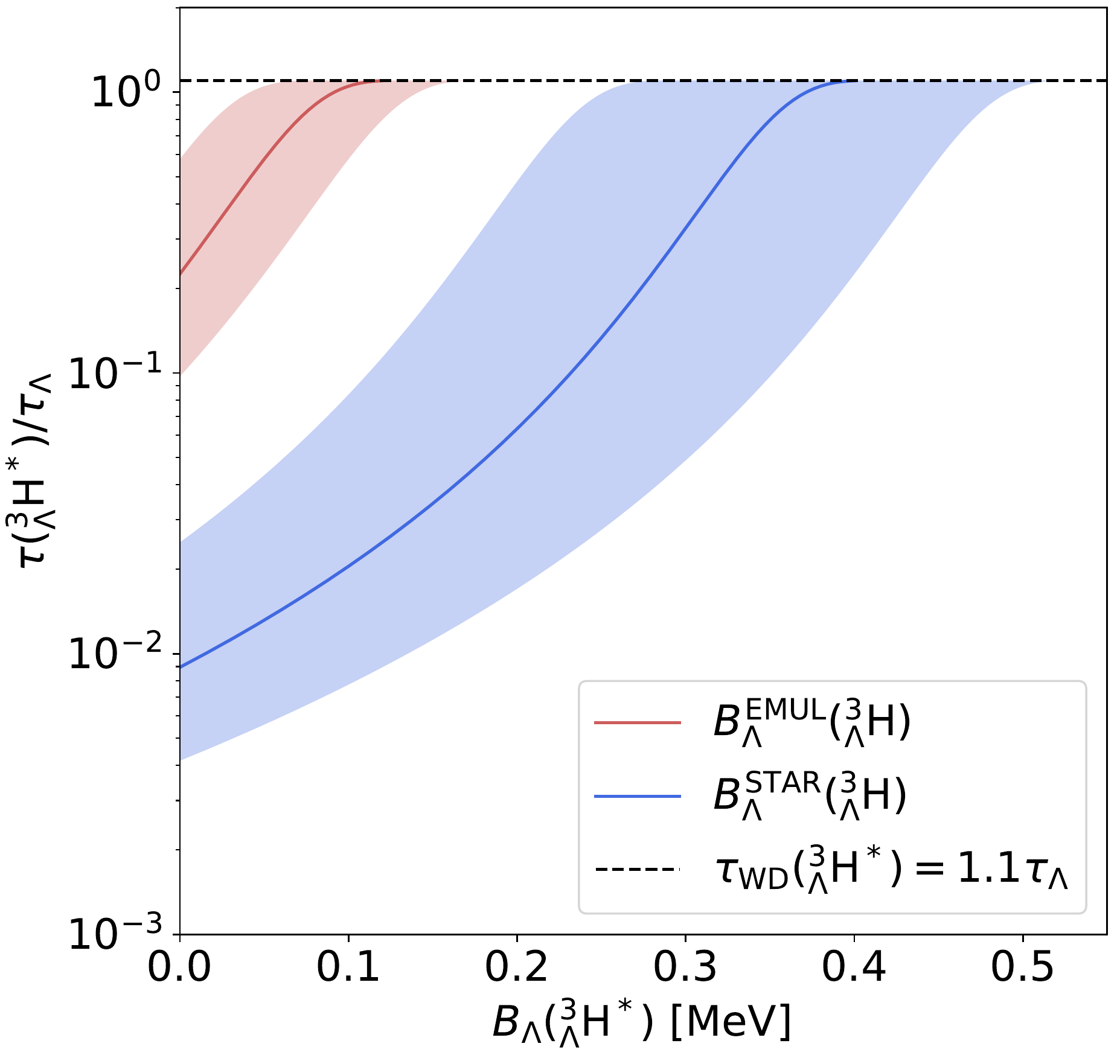}
\caption{Ratio of the hypertriton excited state lifetime to the free $\Lambda$ lifetime,  $\tau({\frac{3}{2}}^+)/\tau_{\Lambda}$, as a function of $B_\Lambda({\rm ^3_\Lambda H^*})$, calculated for $B_\Lambda^{\rm EMUL}({\rm ^3_\Lambda H})$) (red) and
$B_\Lambda^{\rm STAR}({\rm ^3_\Lambda H})$ (blue).  
Shaded areas show the uncertainty induced by the experimental error in the values of $B_\Lambda({\rm ^3_\Lambda H})$.}
\label{3LH32B:3}
\end{figure}


In Fig.~\ref{3LH32B:3} we show $\tau({\frac{3}{2}}^+)/\tau_{\Lambda}$ as a function of $B_\Lambda({\rm ^3_\Lambda H^*})$ for both values 
$B_\Lambda^{\rm STAR}({\rm ^3_\Lambda H})$ and $B_\Lambda^{\rm EMUL}({\rm ^3_\Lambda H})$,  
assuming a fixed value $\tau_{\rm WD} = 1.1\tau_{\Lambda}$. The figure demonstrates strong sensitivity of the hypertriton excited state lifetime  on its binding energy, as well as the energy splitting between the $\rm ^3_\Lambda H^*(3/2^+)$ state and the $\rm ^3_\Lambda H(1/2^+)$ ground state.
Such behavior is due to the strong $\Delta E$ dependence of the M1 deexcitation. It is to be noted that we considered $\tau_{\rm WD}$ fixed though it is energy dependent as well (see ref.~\cite{pgfg20} for the energy dependence of the hypertriton ground-state lifetime). However, the energy dependence of $\tau_{\rm WD}$ is considerably weaker than that of $\tau_{\rm M1}$ and could thus be neglected. Apparently, with increasing $\Delta E$ radiative M1 decay begins to dominate and in certain case [$B_\Lambda^{\rm STAR}({\rm ^3_\Lambda H})$ and just bound $\rm ^3_\Lambda H^*$] corresponding $\tau_{\rm M1}$ lifetime might be up to two orders of magnitude shorter than $\tau_{\rm WD}$. Consequently, in order to search for possibly bound hypertriton excited state one has to resort to both weak decay measurements and $\gamma$ spectroscopy.
\section{Summary}
In the present paper we have explored to what extent the empirical input, represented by the value of the $\Lambda$ separation energy of the hypertriton $^3_{\Lambda}{\rm H}(1/2^+)$ (and also the $\Lambda N$ interaction at threshold), affects calculated binding energies of the hypernuclear trios $\Lambda nn$ and $^3_{\Lambda}{\rm H^*}(3/2^+)$, as well as the 5-body system $^5_{\Lambda}$He. This study has been stimulated by the recent measurement of the STAR collaboration, claiming more tightly bound hypertriton than considered so far, relatively large experimental errors of $B_{\Lambda}({\rm ^3_\Lambda H})$, as well as the lack of $\Lambda N$ scattering data and consequent rather large uncertainty in the values of $\Lambda N$ scattering lengths. Implications of increasing $B_{\Lambda}({\rm ^3_\Lambda H})$ from the emulsion value to the STAR value have been considered recently in the $\chi$EFT work of Le et al. \cite{le20} who focused primarily on binding energies of $\rm ^4_\Lambda He$ and $\rm ^7_\Lambda Li$.

Our calculations have been performed within a LO \nopieft with two- and three-body contact terms. The $\Lambda N$ LECs were fixed by the spin-singlet and spin-triplet $\Lambda N$ scattering lengths given by various interaction models. The experimental values of the $\Lambda$ separation energy $B_\Lambda ({\rm ^4_\Lambda H};0^+)$, excitation energy $E_{\rm exc} ({\rm ^4_\Lambda H^*};1^+)$, and either $B^{\rm EMUL}_{\Lambda}({\rm ^3_\Lambda H};1/2^+)$ or $B^{\rm STAR}_{\Lambda}({\rm ^3_\Lambda H};1/2^+)$ served as a constraints to fix $\Lambda NN$ LECS. Hypernuclear bound states were calculated using SVM; location of the continuum states poles were determined by the IACCC method.  

We first explored the case of $^5_{\Lambda}$He and found that even for rather different strengths of the $\Lambda N$ spin-singlet and spin-triplet interactions the calculated 
$\Lambda$ separation energy $B_{\Lambda}({\rm ^5_\Lambda He})$ changes only moderately. We then considered a much wider range of the $\Lambda N$ scattering lengths constraints and demonstrated that  $B_{\Lambda}({\rm ^5_\Lambda He})$ imposes rather strict limitations on the $\Lambda N$ spin-triplet channel. 
Moreover, both values $B^{\rm EMUL}_{\Lambda}({\rm ^3_\Lambda H})$ and $B^{\rm STAR}_{\Lambda}({\rm ^3_\Lambda H})$  were found acceptable within the LO truncation error as they both could lead to $B_{\Lambda}({\rm ^5_\Lambda He})$ in accord with experiment. This result follows from the possibility to fit three given $s$-shell binding energies, those of $\rm ^3_\Lambda H$, $\rm ^4_\Lambda H$ and $\rm ^4_\Lambda H^*$ for chosen values of $\Lambda N$ scattering lengths (2-body LECs), at the expense of three 3-body LECs within LO \nopieft calculations.


Careful examination of the $\Lambda nn$ pole position,  including all experimental errors in few body inputs taken into account, revealed that chances to get a narrow resonant or even bound $\Lambda nn$ state further decrease with increasing  $B_{\Lambda}({\rm ^3_\Lambda H})$. We thus deem a direct observation of the $\Lambda nn$ system in experiment even more unlikely. 

The excited state of the hypertriton $^3_{\Lambda}{\rm H^*}(3/2^+)$ is found predominantly as a virtual state whose position moves towards the bound state region with increasing $B_{\Lambda}({\rm ^3_\Lambda H})$, eventually converting to a weakly bound state for some $\Lambda N$ interactions strengths. The $^3_{\Lambda}{\rm H^*}$  pole close to the $\Lambda d$ threshold would manifest itself in increased $\Lambda d$ elastic scattering cross-section $\sigma_{\Lambda d}$. Unfortunately, 
$\Lambda d$ scattering experiments are not anticipated due to their extreme difficulty. Nonetheless, even limited experimental information on $\sigma_{\Lambda d}$ at low energies would provide unique probe of the  $\Lambda N$ interaction near threshold.  

The low energy $\Lambda d$ interaction could be experimentally explored by a direct measurement of the corresponding correlation functions. We evaluated the spin-averaged s-wave correlation function within the Lednicky-Lyuboshits approach and demonstrated that its magnitude at low momenta is strongly affected by the 
distance of both states of the hypertriton, the ground state $1/2^+$ and excited state $3/2^+$, from the $\Lambda d$ threshold. However, large uncertainties 
involved in the present calculation do not allow to discriminate between $B^{\rm EMUL}_{\Lambda}({\rm ^3_\Lambda H})$ and $B^{\rm STAR}_{\Lambda}({\rm ^3_\Lambda H})$. Anticipated measurements of 
the $\Lambda d$ correlation function at ALICE@CERN and new experiments at MAMI~\cite{mami18} and  J-PARC, JLAB and ELPH~\cite{nf19} intending to obtain more precise determination of  $B_{\Lambda}({\rm ^3_\Lambda H})$, could help to assess the position of the excited state $^3_{\Lambda}{\rm H^*}$. 

Finally, we elaborated on the eventuality of a weakly bound hypertriton excited state $^3_{\Lambda}{\rm H^*}$. Fitting LECs to a fixed value of $B_{\Lambda}({\rm ^3_\Lambda H^*})$ and four-body constraints while varying $a^{\Lambda N}$ scattering lengths in a wide range
$-3 < a^{\Lambda N} < 0$, we found solutions for $|a^{\Lambda N}_1|<|a^{\Lambda N}_0|$ when $^3_{\Lambda}{\rm H^*}$ is bound and 
the remaining $s$-shell hypernuclei are described in agreement with experiment.  The lifetime of the hypothetically bound excited state of the hypertriton is given by weak decay and the electromagnetic M1 dipole transition to the ground state, which depends strongly on the energy splitting between the two states $3/2^+$ and $1/2^+$. The $\gamma$-ray spectroscopy measurements of the hypertriton planned at the J-PARC K1.1 beamline~\cite{toy} could also help to resolve the question about the nature of the excited state $^3_{\Lambda}{\rm H^*}(3/2^+)$ and determine its binding energy if it is bound.   

\begin{acknowledgments}
This work was partly supported by the Czech Science Foundation GACR grant 19-19640S.
The work of MS and NB was supported by the Pazy Foundation and by the Israel Science Foundation grant 1308/16. 
Furthermore, the work of AG and NB is part of a project funded by the European Union's Horizon 2020 research and innovation
programme under grant agreement No. 824093.
\end{acknowledgments}

\appendix*

\section{Two-body and three-body energy contributions to hypernuclear binding energies}
\label{sec:app} 

This Appendix provides details of some results obtained in \nopieft(LO) 
$s$-shell hypernuclear calculations initiated in Ref.~\cite{CBG18} for bound states 
and pursued in the present work to study unbound systems. In these 
calculations the $NN$ and $\Lambda N$ two-body LECs are determined from the 
corresponding scattering lengths input. Fitting to $B({^3}{\rm H})$=8.482~MeV, 
a three-body $NNN$ LEC with $I=S=1/2$ is determined. Applying the 
$\chi$EFT(LO) model \cite{lo06} scattering lengths, for example (see Fig.~\ref{interactions} here), 
and using ${_{\Lambda}^3}{\rm H}$, ${_{\Lambda}^4}{\rm H}(0^+)$ 
and ${_{\Lambda}^4}{\rm H}^*(1^+)$ $B_{\Lambda}$ input values, 
the three three-body LECs and the corresponding energy contributions $\langle 
V_{\Lambda NN}^{IS}\rangle$ listed in the tables below are extracted from 
the bound state calculations of these systems. This enables a well-defined 
$^4$He and ${_{\Lambda}^5}{\rm He}$ binding energy calculation, with results 
for ${_{\Lambda}^5}{\rm He}$ listed in the last line of each of the tables 
below

\begin{table}[htb]
\begin{center}
\caption{Potential energy contributions (MeV) in $A$=3,4,5 $s$-shell 
hypernuclei calculated for $\lambda$=1~fm$^{-1}$ using the $\chi$EFT(LO) 
$\Lambda N$ scattering lengths \cite{lo06}.}
\begin{tabular}{lccccccc}
\hline
${_\Lambda^Z}$A & $\langle V_{NN} \rangle$ & $\langle V_{NNN} \rangle$ &
$\langle V_{\Lambda N} \rangle$ & $\langle V_{\Lambda NN}^{0\frac{1}{2}}
\rangle$ & $\langle V_{\Lambda NN}^{1\frac{1}{2}} \rangle$ & $\langle
V_{\Lambda NN}^{0\frac{3}{2}} \rangle$ & $B_{\Lambda}$  \\
\hline 
$^3_\Lambda$H   &       -11.30  &       0.00    &       -1.41   &       -0.32  
 &       -       &       -       &       0.13    \\                            

$^4_\Lambda$H      &       -35.65  &       -0.46   &       -6.40  
 &       -1.98   &       0.39    &       -       &       2.16    \\
$^4_\Lambda$H$^*$      &       -34.34  &       -0.43   &       -4.37  
&       -0.16   &       0.28    &       -0.79   &       1.07    \\              

$^5_\Lambda$He  &       -92.49  &       -2.97   &       -12.07  &       -2.32  
 &       1.57    &       -2.92   &       6.29    \\ 
\hline
\end{tabular}
\label{tab:lam=1}
\end{center}
\end{table}

Tables~\ref{tab:lam=1} and \ref{tab:lam=4} list results obtained by using 
cutoff values $\lambda=1$~fm$^{-1}$ and $\lambda=4$~fm$^{-1}$, respectively. 
Note how the 3-body potential contributions turn from {\it weak attraction} 
for $\lambda=1$~fm$^{-1}$ (except in the $I=1,S=1/2$ channel) into larger-size 
{\it repulsion} for $\lambda=4$~fm$^{-1}$. This $\Lambda NN$ repulsion 
is responsible for eliminating largely the $\sim$3~MeV overbinding of 
$^5_\Lambda$He for $\lambda=1$~fm$^{-1}$, as seen also in Fig.2a, almost 
reproducing for $\lambda=4$~fm$^{-1}$ the actual binding energy value 
$B_{\Lambda}^{\rm exp}({^5_\Lambda}{\rm He})$=3.12~MeV. We note that whereas 
each of the repulsive $\Lambda$ kinetic energy and the 2-body $\Lambda N$ 
attractive potential energy increases steadily with increasing the cutoff 
$\lambda$, the repulsive 3-body $\Lambda NN$ potential contribution remains 
finite and relatively small, less than 10 MeV, upon increasing $\lambda$ (see Fig.~2b in Ref.~\cite{CBG19}).

\begin{table}[htb]
\begin{center}
\caption{Potential energy contributions (MeV) in $A$=3,4,5 $s$-shell 
hypernuclei calculated for $\lambda$=4~fm$^{-1}$ using the $\chi$EFT(LO) 
$\Lambda N$ scattering lengths \cite{lo06}.}
\begin{tabular}{lccccccc}
\hline
${_\Lambda^Z}$A & $\langle V_{NN} \rangle$ & $\langle V_{NNN} \rangle$ &
$\langle V_{\Lambda N} \rangle$ & $\langle V_{\Lambda NN}^{0\frac{1}{2}}
\rangle$ & $\langle V_{\Lambda NN}^{1\frac{1}{2}} \rangle$ & $\langle
V_{\Lambda NN}^{0\frac{3}{2}} \rangle$ & $B_{\Lambda}$  \\
\hline
$^3_\Lambda$H   &       -37.11  &       0.00    &       -5.45   &       0.92   
 &       -       &       -       &       0.13    \\                            
  
$^4_\Lambda$H      &       -88.38  &       8.06    &       -22.71 
 &       3.67    &       2.38    &       -       &       2.16    \\
$^4_\Lambda$H$^*$      &       -87.75  &       8.05    &       -14.99 
 &       0.23    &       1.68    &       2.08    &       1.07    \\

$^5_\Lambda$He  &       -156.47 &       23.89   &       -27.63  &       1.72   
 &       4.05    &       3.81    &       3.06    \\
\hline
\end{tabular}
\label{tab:lam=4}
\end{center}
\end{table}

\end{document}